\documentclass[pra,twocolumn,notitlepage,superscriptaddress,nofootinbib,longbibliography]{revtex4-1} 
        
\usepackage{graphicx}
\usepackage{natbib}
\usepackage{epstopdf}
\usepackage{amsmath}
\usepackage{amssymb}
\usepackage{mathbbol}
\usepackage{xcolor}
\usepackage[colorlinks,citecolor=blue,urlcolor=blue]{hyperref}

\newcommand{\abs}[1]{\left\vert#1\right\vert}
\newcommand{\beq}{\begin{equation}}
\newcommand{\eeq}{\end{equation} \smallskip}
\newcommand{\beqy}{\begin{eqnarray}}
\newcommand{\eeqy}{\end{eqnarray} \smallskip}

\newcommand{\bit}{\begin{itemize}}
\newcommand{\eit}{\end{itemize}}
\newcommand{\bmat}{\begin{pmatrix}}
\newcommand{\emat}{\end{pmatrix}}

\begin{document}

%% title, authors etc.
\title{Tuning across Universalities with a Driven Open Condensate}

\author{A. Zamora}

\affiliation{Department of Physics and Astronomy, University College London,
  Gower Street, London, WC1E 6BT, United Kingdom}

\author{L. M. Sieberer}

\affiliation{Department of Physics, University of California, Berkeley, California 94720, USA
}

\author{K. Dunnett}

\affiliation{Department of Physics and Astronomy, University College London,
  Gower Street, London, WC1E 6BT, United Kingdom}

\author{S. Diehl}

\affiliation{Institute of Theoretical Physics, University of Cologne, D-50937
  Cologne, Germany}

\author{M. H. Szyma\'nska}

\affiliation{Department of Physics and Astronomy, University College London,
Gower Street, London, WC1E 6BT, United Kingdom}
\affiliation{Kavli Institute for Theoretical Physics, University of California,
Santa Barbara, CA 93106-4030, USA}
\email{m.szymanska@ucl.ac.uk}

\date{\today}
 
%\pacs{}

\begin{abstract}
  Driven-dissipative systems in two dimensions can differ substantially from
  their equilibrium counterparts. In particular, a dramatic loss of off-diagonal
  algebraic order and superfluidity has been predicted to occur due to the
  interplay between coherent dynamics and external drive and dissipation in the
  thermodynamic limit. We show here that the order adopted by the system can be
  substantially altered by a simple, experimentally viable, tuning of the
  driving process. More precisely, by considering the long-wavelength phase
  dynamics of a polariton quantum fluid in the optical parametric oscillator
  regime, we demonstrate that simply changing the strength of the pumping
  mechanism in an appropriate parameter range can substantially alter the level
  of effective spatial anisotropy induced by the driving laser, and move
  the system into distinct scaling regimes. These include: (i) the classic
  algebraically ordered superfluid below the Berezinskii-Kosterlitz-Thouless
  (BKT) transition, as in equilibrium; (ii) the non-equilibrium,
  long-wave-length fluctuation dominated Kardar-Parisi-Zhang (KPZ) phase; and
  the two associated topological defect dominated disordered phases caused by
  proliferation of (iii) entropic BKT vortex-antivortex pairs or (iv) repelling
  vortices in the KPZ phase. Further, by analysing the renormalization group
  flow in a finite system, we examine the length scales associated with these
  phases, and assess their observability in current experimental conditions.
\end{abstract}

\maketitle

%%%%%%%%%%%%%%%%%%%%%%%%%%%%%%%%%%%%%%%%%%%%%%%%%%%%%%%%%%%
%%%%%%%%%%%%%%%%%%%%%%%%%%%%%%%%%%%%%%%%%%%%%%%%%%%%%%%%%%%
\section{Introduction}
%%%%%%%%%%%%%%%%%%%%%%%%%%%%%%%%%%%%%%%%%%%%%%%%%%%%%%%%%%%
%%%%%%%%%%%%%%%%%%%%%%%%%%%%%%%%%%%%%%%%%%%%%%%%%%%%%%%%%%%

The concept of universality permits to order and classify a great variety of
different physical systems in terms of their common collective behaviour in the
long-wavelength limit. While dynamical critical phenomena in equilibrium are by
now quite well understood~\cite{hohenberg1977theory}, the extension to
non-equilibrium is a relatively new field. At the same time, due to an
unprecedented experimental progress on a range of light-matter
realisations~\cite{RevModPhys.85.299} in recent years, there is a particular
interest in collective behaviour of driven-dissipative quantum systems.
Despite the fact that energy is not conserved, the detailed balance condition is
broken and fluctuation-dissipation relations are not
satisfied~\cite{Sieberer2015,sieberer2015keldysh}, it has been shown than some
three-dimensional driven-dissipative systems close to a critical point may show
emergent fluctuation-dissipation relations and, therefore, universal asymptotic
thermalisation~\cite{mitra06, dalla10:_quant,
  diehl10:_dynam_phase_trans_instab_open, diehl08:_quant,
  mitra11:_mode_coupl_induc_dissip_therm, mitra12:_therm, oztop12:_excit,
  torre13:_keldy, wouters06:_absen, sieberer2013dynamical,Sieberer2014}. This,
however, was suggested not to be the case for some two dimensional systems,
where the dissipation has a more profound qualitative effect, completely
destroying the analogous equilibrium order, and bringing the system to a
different universality class~\cite{PhysRevX.5.011017}.

This finding was of particular importance in the context of
driven-dissipative two-dimensional bosonic superfluids, such as for
example exciton-polaritons in semiconductor microcavities,
where collective phase fluctuations were found to preclude the algebraic
order in the thermodynamic limit, leading to a stretched exponential decay of
first order coherence characteristic of a Kardar-Parisi-Zhang phase
(KPZ)~\cite{PhysRevX.5.011017}.
Even if later estimates of the KPZ length scales for incoherently driven
  microcavities appeared to be beyond the reach of current experiments, and the
presence of free vortices with screened repulsive
interactions~\cite{Aranson1998} might preclude the possibility of the KPZ
phase~\cite{sieberer2016lattice,wachtel2016electrodynamic}, the emerging order
and the type of phase transition in these systems is still subject to an intense
debate \cite{he2016spacetime}.
This is particularly true in light of the fact that exact stochastic simulations able to account for
vortices~\cite{dagvadorj2015nonequilibrium}, but also experiments~\cite{roumpos2012power},
observed a clear transition from exponential decay of correlations to algebraic
order but with an algebraic exponent $\alpha$ as large as four times the
equilibrium upper bound, when approaching the BKT transition, suggesting an
``over-shaken'' but a superfluid state~\cite{dagvadorj2015nonequilibrium}.

In this work, investigating parametrically driven polaritons, we show that the
type of order adopted is in fact not an intrinsic property of the system but can
be strongly sensitive to the driving process able to tune the system between two
different universality classes by only a relatively small change in the driving
strength. The key feature we exploit here is the spatial anisotropy that is
  imprinted on the system by the wave vector of the driving laser. In the
  long-wavelength theory for parametrically driven polaritons, which we derive in
  Sec.~\ref{sec:syst-theor-descr}, the effective degree of anisotropy is
  measured by a single parameter $\Gamma$. This quantity depends in a
  non-trivial way on the system parameters and, in particular, on the driving
  strength.
In the region close to the optical parametric oscillator (OPO) upper threshold,
but at lower powers, the system develops a steady state, which (if vortices
remain bound) falls into the KPZ universality
class~\cite{PhysRevLett.56.889,PhysRevX.5.011017} with a non-equilibrium fixed
point~\cite{PhysRevX.5.011017,PhysRevLett.111.088701} and no counterpart in
equilibrium systems. However, by increasing the strength of the external drive
towards the OPO upper threshold, the effective anisotropy crosses a critical
value and the properties of the system are governed by an equilibrium fixed
point. The system thus falls into the Edwards-Wilkinson (EW) universality
class~\cite{Edwards17}, which captures the universal properties of many
different equilibrium systems, particularly the low temperature spin-wave theory
of the $XY$ model~\cite{hohenberg1977theory}, exhibiting a BKT transition to
off-diagonal algebraic order ensuring superfluidity.
Note, that the equilibrium fixed point is approached for larger driving
strengths than the non-equilibrium one, suggesting that we are not simply
observing an approach to equilibrium as the external dissipation diminishes, but
rather a more profound interplay between drive, dissipation and spatial
anisotropy. 

The various universal scaling regimes that can be accessed with polaritons
  were first discussed in the context of \emph{incoherently} driven systems in
  Ref.~\cite{PhysRevX.5.011017}. However, while in both driving schemes the
  effective long-wavelength theory takes the form of the (anisotropic) KPZ
  equation, which is the origin of the rich universal behaviour, the underlying
  physics is completely different. In the case of incoherent pumping, the KPZ
  equation follows from a standard hydrodynamic description of the dynamics of
  the polariton fluid, and governs fluctuations of the phase of the
  condensate. On the other hand, coherent laser driving pins the condensate
  phase at the pump wave-vector, and the derivation of the long-wavelength
  theory is much more subtle: here, it is the \emph{relative} phase of signal
  and idler modes, which are both macroscopically populated above the OPO
  threshold, that is free to fluctuate. This is the Goldstone mode that
  determines the physics at large scales. As a consequence of the different
  mechanisms leading to the KPZ equation, the natural scales of the coefficients
  appearing in this equation are vastly different in incoherently and coherently
  pumped polaritons. In particular, in the former case both the characteristic
  KPZ non-linearity and the degree of anisotropy are typically
  small~\cite{PhysRevX.5.011017}, making the observation of novel
  non-equilibrium features challenging. Contrary to that, here we show that in
  OPO systems those quantities can be tuned over a wide range of values simply
  by changing the driving strength. Hence, we determine under which conditions
  these scaling regimes are feasible in current experiments on inorganic and
  organic microcavities. Moreover, we analyse the renormalisation group (RG)
  flow equations in a finite system, and estimate the relevant length scales by
  varying the control parameters such as the external laser drive, and the
  detuning between the cavity photons and the excitons.

The phenomena we discuss here are fundamentally induced by strong
  fluctuations, making the mean field approach, that has been applied in most of
  the existing literature on polaritons in the OPO
  regime~\cite{Whittaker2005,Wouters2007,Marchetti2012}, insufficient. Much
  rather, the RG techniques we use are tailored to address universal behaviour
  that occurs on large length and time scales, beyond what is accessible with
  exact numerical methods~\cite{dagvadorj2015nonequilibrium}.

By establishing the universal regimes that are accessible with parametrically 
  driven polaritons, we take a major step towards 
  understanding of phases and phase transitions in 2D driven-dissipative
  systems. In particular, we show that the universal physics in OPO polaritons
  is much richer than anticipated, and has surprising connections to seemingly
  remote fields such as collective behaviour in active
  systems~\cite{PhysRevLett.111.088701}, thus opening a whole new perspective on
  coherently driven polaritons.

%---- Figure ----------------

\begin{figure}
%\centering
  \includegraphics[width=8cm]{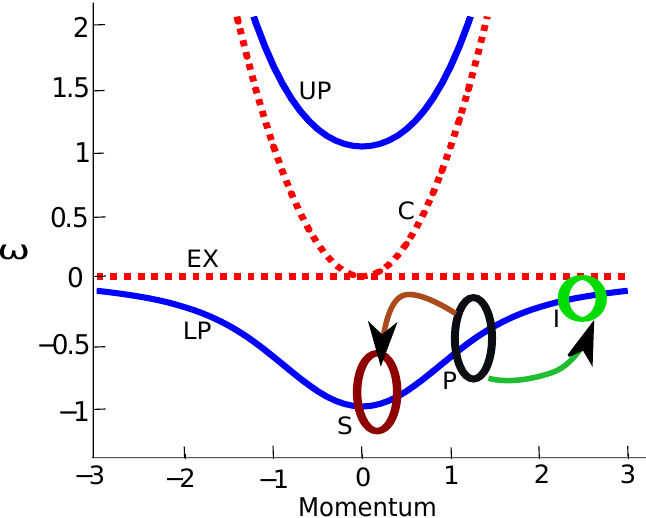}
  \caption{ \textbf{Polariton system.} Dispersions
    for the exciton (EX) and cavity photon (C) with zero detuning
    mixing to form upper (UP) and lower polaritons (LP). The external
    drive introduces pairs of lower polaritons with momentum
    $\mathbf{k}_p$ and energy $\omega_p$ (P) which scatter into the
    signal (S) and idler (I) states while conserving energy and
    momentum.}
\label{fig:dispersion_relation} 
\end{figure}

\begin{figure}
%\centering
    \includegraphics[width=8cm]{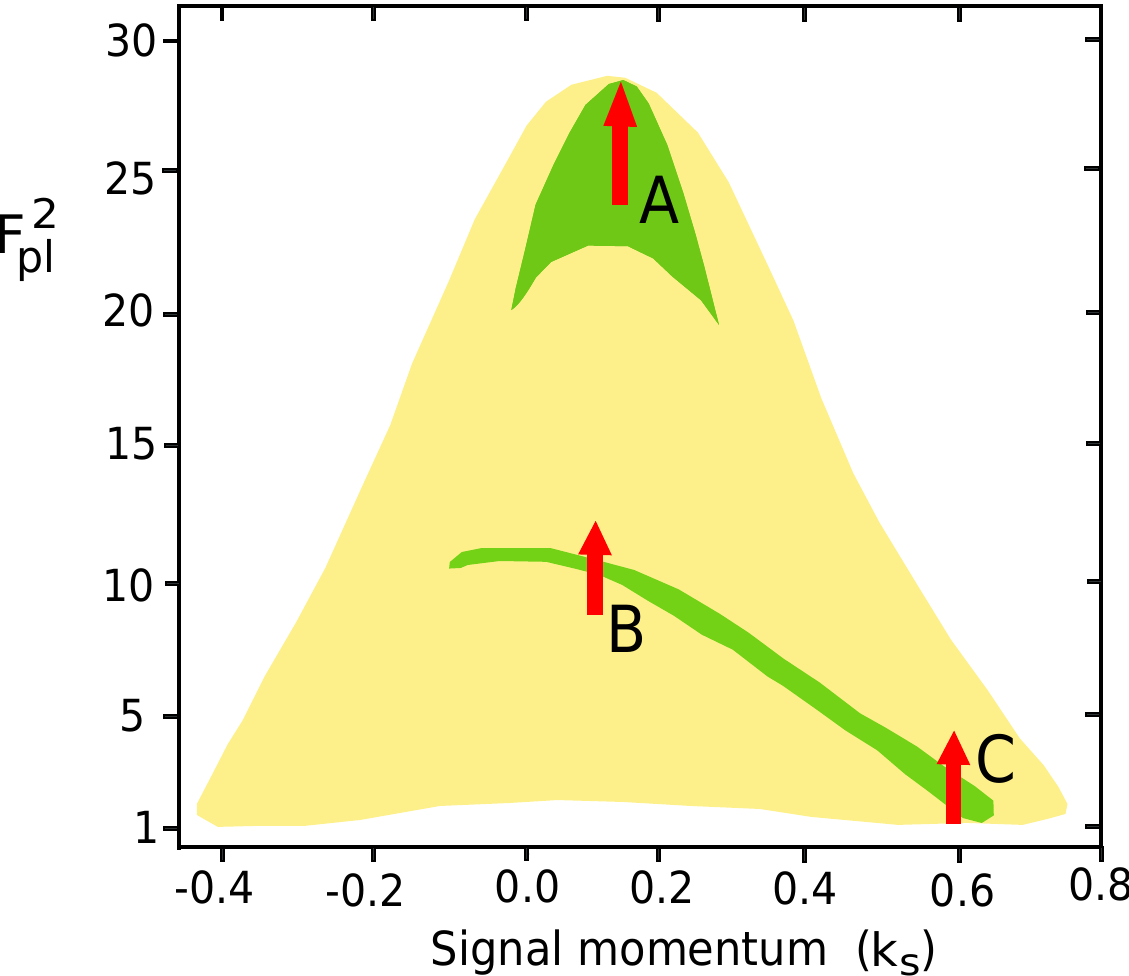}
    \caption{\textbf{Stability diagram.} Regions of instability of the pump-only
      state towards an OPO solution (yellow and green). The simplest three-mode
      OPO state is stable for parameters marked by green, i.e. at $k_s$ around
      0.1 for high $F^2_p$ (region A and B) and over a more extended higher $k_s$
      range at low $F^2_p$ (region C). In the yellow region both the pump-only and
      the three mode ansatz are unstable. The red arrows indicate the direction
      of increasing pump power shown in Figs. \ref{fig:lenghts_galaa_parameters}, \ref{fig:isotropic_phase_non_zero_detuning} and \ref{fig:lengths_intermediate_fp_galaa}.
      Note that the pump intensity
      (vertical axis) is normalized to the lower threshold,
      $F^2_{\mathit{pl}}\equiv (F_p/F^{\textrm{lo}}_p)^2$. Parameters are as in the
      text with zero detuning and $k_p$=1.4.}
\label{fig:mean_field} 
\end{figure}

%---- End Figure ----------------

% %%%%%%%%%%%%%%%%%%%%%%%%%%%%%%%%%%%%%%%%%%%%%%%%%%%%%%%%%%%
% %%%%%%%%%%%%%%%%%%%%%%%%%%%%%%%%%%%%%%%%%%%%%%%%%%%%%%%%%%%
% \section{The system and semiclassical description}
% %%%%%%%%%%%%%%%%%%%%%%%%%%%%%%%%%%%%%%%%%%%%%%%%%%%%%%%%%%%
% %%%%%%%%%%%%%%%%%%%%%%%%%%%%%%%%%%%%%%%%%%%%%%%%%%%%%%%%%%%

\section{System and theoretical description}
\label{sec:syst-theor-descr}

Exciton-polaritons are bosonic quasi-particles emerging in the regime of strong
coupling between excitons in semiconductors and a cavity photon
mode~\cite{RevModPhys.85.299,deng2010exciton} (see
Fig.~\ref{fig:dispersion_relation}). The coherent mixing between light and
matter excitations results in the emergence of two bands in these systems,
termed the upper and lower polariton. Due to mirror imperfections the lifetime
of photons and hence polaritons is finite, necessitating continuous external
laser driving to maintain their finite population. In the stationary
state resulting from the compensation of gain and losses, detailed balance
is violated and the system is therefore not in thermal equilibrium.

The properties of this non-equilibrium stationary state are vitally
influenced by the implementation of the laser driving. In particular,
if the laser frequency is chosen to resonantly populate highly excited
states, the polaritons generated in this way undergo complex
scattering before condensing in the lower polariton
band. All coherence of the exciting laser is lost in these processes
and not transferred to the lower polariton states. The dynamics of the
incipient condensate under such \emph{incoherent} pumping is commonly
described \emph{phenomenologically} in terms of a generalized
Gross-Pitaevskii equation~\cite{RevModPhys.85.299}. In contrast to the
incoherent pumping scheme, in the \emph{coherent} scheme polaritons
are excited with a monochromatic external laser acting resonantly on
or close to the lower polariton
dispersion~\cite{savvidis2000angle}. The absence of complex scattering
processes facilitates an \emph{ab initio} rather than a
phenomenological description as we detail below. As a consequence of
coherent driving of lower polaritons, two symmetries which generically
are present for incoherent pumping should be discussed in some more
detail at this point: (i) The U(1) symmetry under rotations of the
phase of the lower polariton field, and (ii) symmetry under spatial
rotations.

The second point (ii) is addressed by specifying the coherent driving
term for our problem, $f^*_p \psi_p + \mathrm{h.c.}$ with external force $f_p =
F_p e^{i \left( \mathbf{k}_p \cdot \mathbf{r} - \omega_p t \right)}$
and pump field $\psi_p$. Clearly, the directionality imprinted by the
external driving field explicitly breaks the rotational symmetry of
the problem on the microscopic level. It is a key point of this paper
to elaborate on the consequences of this fact for the macroscopic
observables.

Regarding (i), we need to specify the interaction term between the
pump field $\psi_p$, and the signal and idler modes $\psi_s,\psi_i$,
respectively. It describes the interconversion of two pump field
photons into a pair of signal and idler photons (see Fig. 1), $\sim
\psi^*_i\psi^*_s \psi_p^2 + \mathrm{h.c.}$. To begin with, we thus have three
phases of pump, signal and idler fields. However, the external
coherent pump term locks the phase of the pump field $\psi_p$ via the
coherent drive term, in turn locking the \emph{sum} of phases of
signal and idler fields $\psi_s$ and $\psi_i$ via the interaction
term. On the other hand, their \emph{difference} is not fixed by the
dynamics of the system. Thus, there is one remaining U(1) phase
rotation invariance left, which is generated by the transformation
\begin{equation}
  \label{eq:u1_symmetry}
  \begin{pmatrix}
    \psi_s \\
    \psi_i \\
    \psi_p
  \end{pmatrix}
  \mapsto
  \begin{pmatrix}
    e^{i\alpha}    &    0    &   0 \\
    0            & e^{-i\alpha} &  0 \\
    0 & 0 & 1
  \end{pmatrix}
  \begin{pmatrix}
    \psi_s \\
    \psi_i \\
    \psi_p
  \end{pmatrix}.
\end{equation}  
The residual U(1) symmetry introduced in this way can be broken spontaneously,
and is responsible for the existence of a gapless phase mode above the OPO
threshold~\cite{wouters06:_absen,stevenson2000continuous,baumberg2000parametric,tartakovskii2002polariton},
which in turn governs the long distance coherence behaviour of the OPO condensate
to be investigated in this paper.

More precisely, in a coherently driven system, depending on the strength of the external pump
power, we distinguish two different regimes: The \emph{pump-only} state, with
only one mode, $\psi_p$, substantially occupied (white region in
Fig.~\ref{fig:mean_field}). Such a state is characterized by a momentum
$\mathbf{k}_p$ and frequency $\omega_p$, which coincide with the momentum and
frequency of the external pump. The phase of the pump-only state is locked by
the external drive. However, in a certain pumping regime (green region in
Fig.~\ref{fig:mean_field}), pairs of polaritons scatter from the pump state into
two new substantially occupied states, the \emph{signal} $\psi_{s}$ and the
\emph{idler} $\psi_{i}$, with momentum $\mathbf{k}_s$ and $\mathbf{k}_i$ and
frequencies $\omega_s$ and $\omega_i$, respectively (see
Fig.~\ref{fig:dispersion_relation}).
The scattering process is determined by the resonance conditions:
$\mathbf{k}_s+\mathbf{k}_i=2\mathbf{k}_p$ and $\omega_s+\omega_i=2\omega_p$.
Thus, in the OPO regime, the lower polariton field $\psi_{\mathrm{LP}}$ can be
split into three contributions:
\begin{equation}
  \psi_{\mathrm{LP}}(\mathbf{r}, t) = \sum_{j = s, p, i} \psi_j(\mathbf{r},
  t) e^{i \left( \mathbf{k}_j \cdot \mathbf{r} - \omega_j t \right)}.
\label{eq:three_mode}
\end{equation}
In a mean-field treatment ignoring fluctuations, the amplitudes
$\psi_j(\mathbf{r}, t)$ are spatially homogeneous and time-independent. Below we
study the influence of fluctuations on the spatial and temporal coherence
properties of the lower polariton field.

The yellow region in Fig.~\ref{fig:mean_field} marks the range of parameters in
which polaritons are parametrically scattered to more than two additional
momentum states (i.e., there are additional satellite states or the signal mode
extends over a ring in a momentum space). However, here we focus on the regime
in which the three-mode ansatz~\eqref{eq:three_mode} is stable.

As a short digression, it is interesting to note which kind of order is
established upon crossing the OPO threshold. Assuming that the amplitudes in
Eq.~\eqref{eq:three_mode} are constants which we write in density-phase
representation as $\psi_j = \sqrt{\rho_j} e^{i \phi_j}$, the density of lower
polaritons is given by
\begin{multline}
  \label{eq:dw-tc}  
  \abs{\psi_{\mathrm{LP}}(\mathbf{r}, t)} ^2 = \sum_j \rho_j^2 \\ + 2 \left(
    \sqrt{\rho_s \rho_p} \cos(\phi_s - \phi_p + (\mathbf{k}_{si} \cdot \mathbf{r}
    - \omega_{si} t)/2) \right. \\ + \sqrt{\rho_p \rho_i} \cos(\phi_i - \phi_p -
  (\mathbf{k}_{si} \cdot \mathbf{r} - \omega_{si} t)/2) \\ \left. + \sqrt{\rho_s
      \rho_i} \cos(\phi_s - \phi_i + \mathbf{k}_{si} \cdot \mathbf{r} - 
    \omega_{si} t) \right),
\end{multline}
where we denote $\mathbf{k}_{si} = \mathbf{k}_s - \mathbf{k}_i$ and
$\omega_{si} = \omega_s - \omega_i$, and we used the resonance conditions stated
above. Thus, at the mean-field level, the OPO regime is characterized by
density-wave and time-crystal order, with base wave vector and frequency
$\mathbf{k}_{si}/2$ and $\omega_{si}/2$, respectively. The Goldstone mode
alluded to above corresponds to fluctuations of the relative phase
$\phi_s - \phi_i$. These can be incorporated in Eq.~\eqref{eq:dw-tc} by
replacing $\phi_s \to \phi_s + \theta(\mathbf{r}, t)$ and
$\phi_i \to \phi_i - \theta(\mathbf{r}, t)$ (the analysis of fluctuations is
carried out systematically below), which shows that the Goldstone mode is
encoded in the phase of the spatiotemporally periodic order. In particular,
topological defects in $\theta(\mathbf{r}, t)$ are dislocations in the combined
density wave and time crystal. However, by filtering the lower polariton field
in momentum space as is usually done in experiments \cite{sanvitto2010persistent}, it is possible
to single out the signal and idler modes. 
Then, topological defects in
$\theta(\mathbf{r}, t)$ appear as ordinary vortices. Using this approach,
coherence of the OPO state can be quantified by measuring the two-point
correlation function of the signal mode,
\begin{equation}
  \label{eq:g1}
  g^{(1)}(\mathbf{r}, t) = \langle \psi_s(\mathbf{r}, t) \psi_s^{*}(0, 0)
  \rangle.
\end{equation}
In the following, we investigate how mean-field order is affected by
fluctuations of the Goldstone mode.

%%%%%%%%%%%%%%%%%%%%%%%%%%%%%%%%%%%%%%%%%%%%%%%%%%%%%%%%%%%%%%%%%%%
\subsection{Keldysh field integral approach}
\label{sec:keldysh-field-integr}
%%%%%%%%%%%%%%%%%%%%%%%%%%%%%%%%%%%%%%%%%%%%%%%%%%%%%%%%%%%%%%%%%%%

As pointed out above, for coherently pumped polaritons it is possible to derive
a microscopic description. A convenient framework is provided by the Keldysh
field integral formalism (see~\cite{sieberer2015keldysh} for a recent review on
applications to driven-dissipative systems). In this formalism, the coupled
dynamics of excitons and photons under the influence of laser driving and cavity
losses is encoded in a field integral ~\cite{PhysRevB.93.195306}
$Z = \int \mathcal{D}[\psi_{s},\psi_{i},\psi_{p}, \bar{\psi}_s, \bar{\psi}_i,
\bar{\psi}_p] \, e^{i S_{\mathrm{OPO}}}$,
which we write here already in terms of the three polariton modes introduced in
Eq.~\eqref{eq:three_mode}.
Since we are interested in the long range dynamics of the system, we neglect quartic terms involving more than a single quantum field. This \emph{semiclassical} approximation is
applicable in the long-wavelength limit and can be justified formally by
canonical power counting, which shows that the neglected terms are irrelevant in
the renormalisation group sense~\cite{keeling2016superfluidity}. Thus, the action $S_{\mathrm{OPO}}$ becomes $S_C$:
\begin{multline}
 \label{eq:semicl_action}
 S_{C} = \int dt d^2\textbf{r}\Biggl\{
 -(f^*_p\psi^Q_p+f_p\bar{\psi}^Q_p) \; +  \\
 \sum_{j=s,p,i}
 \Biggl[
 \frac{1}{X_j^2}
 (\bar{\psi}^C_j\bar{\psi}^Q_j)
  \begin{pmatrix}
    0            &  [D^A_0]_j^{-1} \\
    [D^R_0]_j^{-1} &  [D^{-1}_0]_j^K \\
  \end{pmatrix}
  \begin{pmatrix}
    \psi^C_j \\
    \psi^Q_j\\
  \end{pmatrix}
  \\
  -g_X\left( \left(2( |\psi^C_s|^2 + |\psi^C_i|^2 + |\psi^C_p|^2) -
      |\psi^C_j|^2\right) \psi^C_j\bar{\psi}^Q_j + \textrm{c.c.}\right) \Biggl]
  \\
  -g_X\left( 2\psi^C_s\psi^C_i\bar{\psi}^C_p\bar{\psi}^Q_p
    +(\psi^C_p)^2(\bar{\psi}^C_i\bar{\psi}^Q_s+\bar{\psi}^C_s\bar{\psi}^Q_i) +
    \textrm{c.c.}  \right) \Biggl\},
\end{multline}
where $\psi_{c, j}$ ($\psi_{q, j}$) is the classical (quantum) component of the
field $\psi_j$. The inverse advanced Green's function reads
$[D^A_0]_j^{-1} = i \partial_t + \omega_j - \omega_{\mathrm{LP}}(\mathbf{k}_j -i
\nabla)-i \gamma_j$,
from which the inverse retarded Green's function can be obtained by using the
relation $[D^R_0]_j^{-1}=([D^A_0]_j^{-1})^\dag$; $\gamma_j$ is the decay rate of
the mode $j$, and the lower polariton dispersion, in dimensionless
units,\footnote{Here and in the following we are using dimensionless units,
  measuring time, length, and energy in $2/\Omega_R$,
  $\sqrt{\hbar/(\Omega_Rm_C)}$ and $\hbar\Omega_R/2$ respectively, where
  $\Omega_R$ is the Rabi frequency of the exciton-photon coupling, and $m_C$ is
  the cavity photon effective mass.} is given by
\begin{equation}
\omega_{\mathrm{LP}} (\mathbf{q})=
\frac{1}{2}\left(
q^2 + \delta_{\mathit{CX}} - \sqrt{ \left(q^2 + \delta_{\mathit{CX}} \right)^2 + 4  }  
\right),
\label{eq:dispersion_lower_pol}
\end{equation}
%
% end for blue!!
where $\delta_{\mathit{CX}}$ is the detuning between cavity photons and
excitons. A typical dispersion relation for zero detuning is shown in
Fig.~\ref{fig:dispersion_relation}. The Keldysh part of the inverse Green's function,
$[D^{-1}_0]_j^K = i 2 \gamma_j$, stems from integrating out the bosonic decay
bath fields in the Markovian approximation~\cite{PhysRevB.93.195306}.
$X_j\equiv X(\mathbf{k}_j)$ is the excitonic Hopfield coefficient of the mode
$j$ with momentum $\mathbf{k}_j$~\cite{deng2010exciton}, and $g_X$ is the
strength of the exciton-exciton interaction. Finally, the monochromatic external
pump is, as mentioned previously, of the form
$f_p = F_p e^{i \left( \mathbf{k}_p \cdot \mathbf{r} - \omega_p t \right)}$, where $F_p$ is taken to be a positive real number.
Since the action $S_C$ is only quadratic in quantum
fields we can make use of Martin-Siggia-Rose formalism
~\cite{keeling2016superfluidity,sieberer2015thesis,altland2010condensed}
and map the functional integral to the set of coupled stochastic differential equations for the
signal, idler, and pump modes, which determine the dynamics of the system:
\begin{equation}
\begin{split}
  i\partial_t\psi_s & = \Omega_s\psi_s-i\gamma_s + \tilde{g}_X \psi^2_p\psi^*_i +
  \xi_s,
  \\
  i\partial_t\psi_i & = \Omega_i\psi_i-i\gamma_i + \tilde{g}_X \psi^2_p\psi^*_s +
  \xi_i,
  \\
  i\partial_t\psi_p & = \Omega_p\psi_p -i\gamma_p+
  2\tilde{g}_X \psi_s\psi_i\psi^*_p + f_p + \xi_p,
\end{split}
\label{eq:SCGPE}
\end{equation}
where $\tilde{g}_X \equiv g_X X^2_pX_iX_s$, $\psi_j\equiv\psi_{j}/X_j$, we use the shorthand notation~\cite{PhysRevB.93.195306}
$\Omega_j \equiv -\omega_j+\omega_{\mathrm{LP}} (\mathbf{k}_j-i\nabla)+
g_X X^2_j\left(2(\tilde{n}_s+\tilde{n}_p+\tilde{n}_i)-\tilde{n}_j\right)$
(with $\tilde{n}_j=X^2_j \abs{\psi_j}^2$) and we redefine the external pump
$X_p f_p \to f_p$. The terms $\xi_{s, i, p}$ are Gaussian noise sources which
have vanishing expectation value, $\langle \xi_j(\mathbf{r},t)\rangle =0$, and
white spectrum,
$ \langle \xi_j(\mathbf{r}, t) \xi^*_{j'}(\mathbf{r}', t')\rangle= 2 \gamma_j
\delta(\mathbf{r}-\mathbf{r}') \delta(t-t') \delta_{jj'}$.
Note that the dynamical equations \eqref{eq:SCGPE} are invariant under the U(1) transformation expressed in \eqref{eq:u1_symmetry}.

% %%%%%%%%%%%%%%%%%%%%%%%%%%%%%%%%%%%%%%%%%%%%%%%%%%%%%%%%%%%
% %%%%%%%%%%%%%%%%%%%%%%%%%%%%%%%%%%%%%%%%%%%%%%%%%%%%%%%%%%%
% \section{Long-wavelength description in the OPO regime}
% %%%%%%%%%%%%%%%%%%%%%%%%%%%%%%%%%%%%%%%%%%%%%%%%%%%%%%%%%%%
% %%%%%%%%%%%%%%%%%%%%%%%%%%%%%%%%%%%%%%%%%%%%%%%%%%%%%%%%%%%
% \label{sec:kpz_equation}

\subsection{Long-wavelength theory in the OPO regime: mapping to the anisotropic
  KPZ equation}
\label{sec:long-wavel-theory}

The stochastic equations~\eqref{eq:SCGPE} for the signal, idler, and pump modes
provide a convenient starting point for deriving the effective long wavelength
theory for polaritons in the OPO regime. We follow the usual strategy of
parametrizing fluctuations around the mean-field solution in the density-phase
representation, i.e., we write the three modes as
\begin{equation}
  \psi_j(\mathbf{r}, t) = \left( \sqrt{\rho_j} + \pi_j(\mathbf{r}, t) \right)
  e^{i \left( \phi_j + \theta_j(\mathbf{r},t) \right)}, 
\label{eq:phase_amplitude}
\end{equation}
where $\sqrt{\rho_j}$ and $\phi_j$ are the homogeneous and stationary mean-field
density and phase, respectively, obtained by solving Eq.~\eqref{eq:SCGPE} with
$\xi_{s, i, p} \equiv 0$; fluctuations around the mean-field solution are
encoded in the fields $\pi_j(\mathbf{r},t)$ and $\theta_j(\mathbf{r},t)$. 
The key point that allows us to considerably simplify the equations resulting from
inserting the ansatz~\eqref{eq:phase_amplitude} in Eq.~\eqref{eq:SCGPE} is that
fluctuations of the relative phase
$\theta(\mathbf{r}, t) = \theta_s(\mathbf{r}, t) - \theta_i(\mathbf{r}, t)$ of
signal and idler modes, i.e., fluctuations of the Goldstone mode, are
\emph{gapless}, due to the U(1) symmetry expressed at \eqref{eq:u1_symmetry}, while fluctuations of all other are \emph{gapped}. 
This implies that in the limit of long wavelength and low frequencies the latter
fluctuations are small and the equations of motion can be linearised in these
variables, which can then be eliminated. Details of this calculation are
given in Appendix~\ref{sec:appendix_kpz}. It results in a single stochastic
equation for the Goldstone mode $\theta$, which takes the form of the
anisotropic KPZ (aKPZ)~\cite{wolf1991kinetic, PhysRevLett.111.088701} equation
with an additional drift term proportional to $\nabla\theta$:
\begin{equation}
  \partial_t \theta = \sum_{i = x, y} \left[ D_i \partial_i^2 \theta
    + \frac{\lambda_i}{2} \left( \partial_i \theta \right)^2 \right] 
  + \mathbf{B} \cdot \nabla \theta + \eta.
\label{eq:akpz}
\end{equation}
The coefficients $D_{x, y}, \lambda_{x, y}$ and $\mathbf{B}$ result from linear
combinations of different parameters appearing in Eqs.~\eqref{eq:SCGPE} (see
Appendix~\ref{sec:appendix_kpz} for details). The Gaussian white noise term
$\eta$ derives from the different noise terms $\xi_j$ and satisfies
$ \langle \eta(\mathbf{r},t)\rangle =0$ and
$ \langle \eta(\mathbf{r}, t) \eta(\mathbf{r}', t')\rangle = 2 \Delta
\delta(\mathbf{r} - \mathbf{r}') \delta(t-t')$,
where the noise strength $\Delta$ is related to the decay rates
$\gamma_j$. In addition to eliminating massive fluctuations as discussed above,
to obtain Eq.~\eqref{eq:akpz} we also expanded the lower polariton
dispersion~\eqref{eq:dispersion_lower_pol} around each mode $j$ with momentum
$\mathbf{k}_j$ to second order in the gradient:
\begin{equation}
  \omega_{\mathrm{LP}}(\mathbf{k}_j - i \nabla) \approx
  \omega_{\mathrm{LP}}(\mathbf{k}_j) - i \omega_{1j} \cdot \nabla - \nabla^T \omega_{2j} \nabla.
\label{eq:expansion_dispersion}
\end{equation}
This expression shows clearly that the finite value of the pump wave vector
$\mathbf{k}_p$ (and hence of the idler wave vector, and in some cases also of
the signal wave vector) lies at the heart of the spatial anisotropy of the
system, which leads in particular to the matrix $\omega_{2j}$ having two
distinct eigenvalues, resulting in $D_x \neq D_y$ and $\lambda_x \neq \lambda_y$
in the effective long-wavelength description Eq.~\eqref{eq:akpz}. This should be
compared to incoherently pumped polaritons, for which the expansion of the lower
polariton dispersion around zero momentum,
$\omega_{\mathrm{LP}}(- i \nabla) \approx \omega_{\mathrm{LP}}(0) - \nabla^2/(2
m_{\mathrm{LP}})$,
where $m_{\mathrm{LP}}$ is the mass of lower polaritons, leads to the
\emph{isotropic} KPZ equation~\cite{PhysRevX.5.011017}.\footnote{However, even
  in incoherently pumped polaritons some degree of anisotropy can be induced by
  the crystal structure and the splitting of transverse electric and transverse
  magnetic cavity modes~\cite{RevModPhys.85.299, Shelykh2010}.} Below we show
how the values $D_{x, y}$ and $\lambda_{x, y}$, and hence the effective degree
of anisotropy, depend on system parameters such as pumping and
detuning. Crucially, by tuning these parameters we can access different
universal scaling regimes of Eq.~\eqref{eq:akpz}.

We note that the drift term $\mathbf{B} \cdot \nabla \theta$ can be eliminated
from Eq.~\eqref{eq:akpz} by introducing a new variable
$\theta'(\mathbf{r}, t) = \theta(\mathbf{r} + \mathbf{v}_0 t, t)$, i.e., by
transforming to a frame of reference that moves at a velocity
$\mathbf{v}_0$. For $\mathbf{v}_0 = \mathbf{B}$, the equation of motion of
$\theta'$ is given by the aKPZ equation without the drift term. It is thus
sufficient to consider the latter equation, and transform back to the laboratory
frame of reference only for calculating observables in terms of the original
variable $\theta$.

%%%%%%%%%%%%%%%%%%%%%%%%%%%%%%%%%%%%%%%%%%%%%%%%%%%%%%%%%%%%%%%%%%%
%%%%%%%%%%%%%%%%%%%%%%%%%%%%%%%%%%%%%%%%%%%%%%%%%%%%%%%%%%%%%%%%%%%
\subsection{Scaling regimes of the anisotropic KPZ equation}
\label{sec:scal-regim-anis}
%%%%%%%%%%%%%%%%%%%%%%%%%%%%%%%%%%%%%%%%%%%%%%%%%%%%%%%%%%%%%%%%%%%
%%%%%%%%%%%%%%%%%%%%%%%%%%%%%%%%%%%%%%%%%%%%%%%%%%%%%%%%%%%%%%%%%%%

In the previous section we showed that fluctuations around the three-mode OPO
state~\eqref{eq:three_mode} are governed by the anisotropic KPZ
equation~\eqref{eq:akpz}. What does this mean for the spatial and temporal
coherence of the polariton condensate as measured by the first order coherence
function Eq.~\eqref{eq:g1}? There are three aspects which make the physics of
Eq.~\eqref{eq:akpz} rich but also complex to analyse: (i) spatial anisotropy,
(ii) the non-linear terms with coefficients $\lambda_{x, y}$, and (iii) the
compactness of $\theta$, which implies that this field can contain topological
defects. Approaching the problem analytically, difficulties (ii) and (iii) can
be controlled perturbatively, if both the non-linearities $\lambda_{x, y}$ and
the vortex fugacity $y$, which is a measure of the probability of
vortex-antivortex pairs forming at a microscopic distance, are small
parameters.\footnote{We note that $y$ depends on the physics on short scales
  (information which is not contained in the long-wavelength description
  Eq.~\eqref{eq:akpz}) but could in principle be calculated from the stochastic
  equations~\eqref{eq:SCGPE}.} Then, as we describe in the following, depending
on (i) the strength of anisotropy quantified by the anisotropy parameter
\begin{equation}
  \label{eq:Gamma}
  \Gamma \equiv \lambda_yD_x / (\lambda_x D_y),
\end{equation}
based on the perturbative treatment we expect strikingly different behaviour in
the weakly and strongly anisotropic regimes, characterized by $\Gamma > 0$ and
$\Gamma < 0$, respectively.

To understand why $\Gamma = 0$ separates these regimes, we first note that
  $D_{x,y} > 0$ is required for Eq.~\eqref{eq:akpz} to be dynamically
  stable; Therefore, $\Gamma < 0$ corresponds to $\lambda_x$ and $\lambda_y$
  having opposite signs. If $\lambda_x$ and $\lambda_y$ have the same sign and
  hence $\Gamma > 0$, it does not make a difference whether $\lambda_{x, y}$ are
  both positive or negative. In fact, the former case is related to the latter
  by the transformation $\theta \to - \theta$ in Eq.~\eqref{eq:akpz} (after the
  drift term has been removed as described above). Thus, the physics can change
  \emph{qualitatively} only when $\lambda_{x, y}$ have opposite sign and thus
  $\Gamma < 0$. This is indeed found to be the case in the RG analysis.

In the weakly anisotropic (WA) regime both the non-linear terms
$\lambda_{x, y}$~\cite{wolf1991kinetic, PhysRevLett.111.088701} and the fugacity
$y$~\cite{wachtel2016electrodynamic} are relevant couplings, i.e., they grow
under renormalisation. In the absence of vortices this would imply that the
correlation function $g^{(1)}(\mathbf{r}, t)$ takes the form of a stretched
exponential with KPZ scaling exponents (see Eq.~\eqref{eq:scaling_kpz}
below). This behaviour would be observable on length and time scales greater than
$L_{\mathrm{KPZ}}$ and $t_{\mathrm{KPZ}}$, respectively, which mark the
breakdown of the perturbative treatment in $\lambda_{x, y}$. However, eventually
vortices might unbind at a scale $L_v$ and after a time
$t_v$~\cite{wachtel2016electrodynamic}, leading to exponential decay of
correlations (and a absence of superfluid behaviour) beyond these scales. For a
detailed discussion of the influence of vortices in this regime see Appendix
\ref{sec:appendix_vortex}.

The physics is quite different in the strongly anisotropic (SA) regime: for
$\Gamma < 0$, the non-linearities $\lambda_{x, y}$ are irrelevant and flow to
zero. Then, the linearised version of Eq.~\eqref{eq:akpz} exhibits a BKT
transition driven by the noise strength (which in turn depends on the loss rates
and the external drive, see Appendix~\ref{sec:appendix_kpz}), i.e., at low noise
a superfluid phase with algebraic order is
possible~\cite{PhysRevLett.111.088701, PhysRevX.5.011017} even in thermodynamic limit of an infinite system. 
Intriguingly, OPO polaritons allow to cross the boundary between the WA and SA
regimes. This is shown in the next section. Before that, in the remainder of
this section, we discuss in detail the renormalisation group (RG) flow and the two scaling regimes of the aKPZ equation.

The RG flow of the aKPZ equation in the absence of
vortices was analysed in Refs.~\cite{wolf1991kinetic,
  PhysRevLett.111.088701}. It can be parameterized in terms of only two
independent quantities, the anisotropy parameter $\Gamma$ introduced in
Eq.~\eqref{eq:Gamma}, and the rescaled dimensionless non-linearity $g$ which is
defined as
\begin{equation}
  \label{eq:g}
  g \equiv \lambda_x^2\Delta/(D_x^2\sqrt{D_x D_y}).
\end{equation}
To leading order in $g$, the RG flow equations read:
\begin{equation}
  \begin{split}
    \frac{d g}{d l} & = \frac{g^2}{32 \pi} \left( \Gamma^2 + 4 \Gamma - 1 \right),
    \\ \frac{d \Gamma}{d l} & = \frac{\Gamma g}{32 \pi} \left( 1 - \Gamma^2 \right).
  \end{split}
  \label{eq:rg_flow_eq}
\end{equation}
As described above, depending on the value of $\Gamma$, we distinguish between WA
and SA regimes. In the former case, $\lambda_x$ and $\lambda_y$ have the same
sign (the coefficients $D_{x, y}$ have to be positive to ensure stability).
Then, the non-linearity is marginally relevant, and the RG flow takes the system
to a strong coupling fixed point at $g_{*}$ which is beyond the scope of the
perturbative treatment. Moreover, $\Gamma \to 1$, i.e., at large scales
rotational symmetry is restored\footnote{For $\Gamma = 1$, by rescaling the
  units of length Eq.~\eqref{eq:akpz} can be brought to isotropic form with
  $D_x = D_y$ and $\lambda_x = \lambda_y$.} and thus the system falls into the
usual isotropic KPZ universality class.

On the other hand, in the SA regime with $\Gamma < 0$, which is realized when the
coefficients $\lambda_x$ and $\lambda_y$ have opposite sign, the non-linearity
$g$ is irrelevant and flows to zero. As a consequence, the aKPZ equation becomes
a linear stochastic differential equation, which is governed by an equilibrium
fixed point at $g = 0, \Gamma = -1$, and the system falls into the EW
universality class~\cite{Edwards17}.

Having discussed how the RG flow of the aKPZ equation is structured by
  different fixed points in the WA and SA regimes, it is natural to ask for
  observable consequences of these findings. Universal scaling behaviour leaves
  its mark in the long-time and long-range decay of correlations. Hence, in the
  following we discuss the form of the correlation function~\eqref{eq:g1}
  implied by these results, and which modifications are to be
  expected due to the possible occurrence of vortices.

%%%%%%%%%%%%%%%%%%%%%%%%%%%%%%%%%%%%%%%%%%%%%%%%%%%%%%%%
%%%%%%%%%%%%%%%%%%%%%%%%%%%%%%%%%%%%%%%%%%%%%%%%%%%%%%%%
\subsubsection{Weakly anisotropic regime}
\label{sec:weakly-anis-regime}
%%%%%%%%%%%%%%%%%%%%%%%%%%%%%%%%%%%%%%%%%%%%%%%%%%%%%%%%
%%%%%%%%%%%%%%%%%%%%%%%%%%%%%%%%%%%%%%%%%%%%%%%%%%%%%%%%

In the WA regime, the two point correlation function~\eqref{eq:g1}
$g^{(1)}(\mathbf{r},t) \propto e^{- C(\mathbf{r}, t)/2}$ (this form assumes that
density fluctuations are negligible as compared to fluctuations of the Goldstone
mode, see Sec.~\ref{sec:long-wavel-theory}), where
$C(\mathbf{r}, t) = \langle \left( \theta(\mathbf{r}, t) - \theta(0,0) \right)^2
\rangle$, shows a stretched exponential decay~\cite{frey1994two}:
\begin{equation}
  C(\mathbf{r}, t) \sim
  \tilde{r}^{2\chi} F_{\mathrm{KPZ}}(c_1t/\tilde{r}^z) \sim
\begin{cases}
\tilde{r}^{2\chi} & \textrm{for }  \tilde{r}^z \gg c_1t, \\
t^{2\chi/z} & \textrm{for } c_1t \gg \tilde{r}^z,
\end{cases}
\label{eq:scaling_kpz}
\end{equation}
where $\tilde{r}^2=(x/x_0)^{2}+(y/y_0)^{2}$ encodes the anisotropy of the system and the parameter $c_1$ depends on the microscopic parameters.
The limiting forms follow from the asymptotic behaviour of the scaling function,
$F_{\mathrm{KPZ}}(w) \sim A_1$ for $w \to 0$ and $F_{\mathrm{KPZ}}(w)\sim A_2 w^{2 \chi/z}$ for
$w \to \infty$, with non-universal constants $A_1$ and $A_2$. In two spatial
dimensions, the roughness exponent is $\chi \approx 0.39$ (see
Refs.~\cite{Halpin-Healy2014, Pagnani2015} for recent numerical investigations
of KPZ scaling and~\cite{Canet2010} for a functional RG analysis), and the
dynamical exponent $z$ can be obtained from the exact scaling relation
$\chi + z = 2$.

The scaling form Eq.~\eqref{eq:scaling_kpz} applies to the co-moving reference
frame (see the discussion below Eq.~\eqref{eq:akpz}), in which the drift term
$\mathbf{B} \cdot \nabla \theta$ is absent. We can calculate
$g^{(1)}(\mathbf{r}, t)$ in the original frame simply by replacing
$\mathbf{r} \mapsto \mathbf{r} + \mathbf{B} t$ in Eq.~\eqref{eq:scaling_kpz},
which yields
\begin{equation}
  C(\mathbf{r}, t) \sim
  \abs{\mathbf{r}+\mathbf{B}t}^{2 \chi}
  F_{\mathrm{KPZ}}(c_1t / \abs{\mathbf{r} + \mathbf{B}t}^z).
\label{eq:kpz_scaling_rest_frame}
\end{equation}
From this expression we explore the consequences of a non-vanishing drift term
$\mathbf{B}$ on the correlations of the system for the KPZ scaling. In
particular, we find for spatial correlations at equal times
\begin{equation}
  \label{eq:kpz_spatial}
  C'(\mathbf{r},0) \sim \tilde{r}^{2\chi},
\end{equation}
which coincides with the result for the case of vanishing drift term. However,
temporal correlations are modified:
\begin{equation}
  \label{eq:kpz_temporal}
  C'(0,t) \sim \left( B t \right)^{2\chi}
  F_{\mathrm{KPZ}}(c_1t/(B t)^z) \sim
  \begin{cases}
    t^{2 \chi/z} & \text{for } t \ll \tau_c,\\
    t^{2 \chi} & \text{for } t \gg \tau_c.\\
  \end{cases}
\end{equation}
Hence, the system exhibits two different exponents, depending on the
time scale.  Initially, the correlator $g^{(1)}(0,t)$ shows
stretched exponential decay with exponent $2 \chi/z$ characteristic of
KPZ scaling. At longer times, the drift term causes the exponent to
increase to $2 \chi$, resulting in a faster decay of temporal
correlations. The crossover time $\tau_c$ at which the transition
between the two regimes occurs can be obtained from: $c_1\tau_c \sim
\left(B \tau_c \right)^z$, leading to $\tau_c \sim
(B^z/c_1)^{1/(1-z)}$.  
The scaling forms~\eqref{eq:kpz_spatial} and~\eqref{eq:kpz_temporal} are
approached on certain length and time scales. For small KPZ non-linearity $g$, the
scale $L_{\mathrm{KPZ}}$ above which spatial correlations are expected to behave
as~\eqref{eq:kpz_spatial} can in the isotropic case be estimated as \cite{PhysRevX.5.011017}
\begin{equation}
  L_{\mathrm{KPZ}} = \xi_0 e^{8 \pi/g},
  \label{eq:length_KPZ}
\end{equation}
and the corresponding time scale, after which scaling behaviour according to
Eq.~\eqref{eq:kpz_temporal} sets in, follows from diffusive scaling and is given
by~\cite{wachtel2016electrodynamic}
$t_{\mathrm{KPZ}} = L_{\mathrm{KPZ}}^2/\bar{D}$.\footnote{While these estimates were
originally derived for isotropic systems, we expect them to remain valid in the
WA regime and adapt the expression for $t_{\mathrm{KPZ}}$ to the latter by
replacing the isotropic diffusion constant by the geometric mean
$\bar{D} = \sqrt{D_x D_y}$.} In Eq.~\eqref{eq:length_KPZ},
$\xi_0 = \hbar/\sqrt{ 2 m_{\mathrm{LP}} g_X \sqrt{n_s n_i}}$ is the healing
length of the system~\cite{marchetti2010spontaneous}.

Finally, as we mentioned previously, taking into account the
compactness of the phase in the KPZ equation in the WA regime,
vortices have been predicted to unbind at a scale $L_v$
\cite{wachtel2016electrodynamic}
\begin{equation}
  L_v = \bar{\xi}_0 e^{2 \bar{D}/\bar{\lambda}},
  \label{eq:vortex_scale}
\end{equation}
leading to exponential decay of correlations beyond.\footnote{As in the above
  estimates of the characteristic KPZ scales we expect to obtain a valid
  estimate throughout the WA regime by replacing the isotropic diffusion
  constant and non-linearity by the geometric averages $\bar{D}$ and
  $\bar{\lambda} = \sqrt{\lambda_x \lambda_y}$. Note that $\bar{\xi}_0$ is
  related to the vortex mobility and may differ substantially from $\xi_0$, see
  Appendix \ref{sec:appendix_vortex}.  }
Thus the
algebraic or KPZ orders in the WA regime might appear only as a finite
size or transient phenomena. We refer the reader to Appendix
\ref{sec:appendix_vortex} for a detailed description of the physics of
the vortices in the WA regime of the compact KPZ equation.

%%%%%%%%%%%%%%%%%%%%%%%%%%%%%%%%%%%%%%%%%%%%%%%%%%%%%%%%%%%%%%
%%%%%%%%%%%%%%%%%%%%%%%%%%%%%%%%%%%%%%%%%%%%%%%%%%%%%%%%%%%%%%
\subsubsection{Strongly anisotropic regime}
\label{sec:strongly-anis-regime}
%%%%%%%%%%%%%%%%%%%%%%%%%%%%%%%%%%%%%%%%%%%%%%%%%%%%%%%%%%%%%%
%%%%%%%%%%%%%%%%%%%%%%%%%%%%%%%%%%%%%%%%%%%%%%%%%%%%%%%%%%%%%%

In the SA regime the RG flow equations~\eqref{eq:rg_flow_eq} approach
the fixed point at $g = 0, \Gamma = -1$, belonging to the EW
universality class~\cite{Edwards17}. Then, for a zero drift term
  $\mathbf{B}$, the correlations decay as power laws both in space
and time~\cite{rutenberg1995phase,szymanska2006nonequilibrium,szymanska2007mean}:
\begin{equation}
  g^{(1)}(\mathbf{r},t) \sim
  \tilde{r}^{-\alpha} F_{\mathrm{EW}}(\tilde{r}^{z'}/(c_2t)) \sim
\begin{cases}
\tilde{r}^{-\alpha} & \text{for }  \tilde{r}^{z'} \gg c_2t, \\
t^{-\alpha/z'} & \text{for }  c_2t \gg \tilde{r}^{z'} ,
\end{cases}
\label{eq:scaling_Xy}
\end{equation}
where the parameter $c_2$ depends on microscopic parameters. The
scaling function $F_{\mathrm{EW}}(w)$ behaves asymptotically as
$F_{\mathrm{EW}}(w) \sim A_1'$ for $w\to \infty$ and as
$F_{\mathrm{EW}}(w) \sim A_2' w^{\alpha/z'}$ for $w\to 0$, where
$A_1'$ and $A_2'$ are non-universal constants. The exponents are $z' =
2$ and $\alpha = \kappa(\infty)/(4\pi)$, with the renormalized scaled
noise
\begin{equation} 
  \kappa(l) = \Delta(l)/\sqrt{D_x(l) D_y(l)},
  \label{eq:ren_scaled_noise}
\end{equation} 
evaluated from the RG flow equations for the aKPZ equation in the limit
$l \to \infty$~\cite{PhysRevLett.111.088701}. 
We obtain the correlations in the original frame of reference by reverting the coordinate
transformation from the co-moving frame in which the drift term in
Eq.~\eqref{eq:akpz} is absent. Replacing
$\mathbf{r} \mapsto \mathbf{r} + \mathbf{B} t$ in Eq.~\eqref{eq:scaling_Xy},
yields
\begin{equation}
  g'^{(1)}(\mathbf{r}, t) \sim
  \abs{\mathbf{r}+\mathbf{B}t}^{-\alpha}
  F_{\mathrm{EW}}(\abs{\mathbf{r} + \mathbf{B}t}^{z'}/(c_2t)).
\end{equation}
We examine the consequences of a non-vanishing drift term
$\mathbf{B}$ on the correlations of the system. In particular, the spatial correlations at equal times behave as
\begin{equation}
  g'^{(1)}(\mathbf{r},0) \sim \tilde{r}^{-\alpha},
\end{equation}
which coincides with the result for the case of vanishing drift term.
However, as in the KPZ scaling regime, temporal correlations are modified:
\begin{equation}
  g'^{(1)}(0,t) \sim \left( B t \right)^{-\alpha}
  F_{\mathrm{EW}}((B t)^{z'}/(c_2t)) \sim
  \begin{cases}
    t^{-\frac{\alpha}{{z'}}} \;\text{for } t \ll \tau'_c,\\
    t^{-\alpha} \; \text{for } t \gg \tau'_c.\\
  \end{cases}
\end{equation}
Thus, as in the WA regime, the system shows two different exponents, depending on the time scale. 
Initially, the correlator $g^{(1)}(0,t)$ shows algebraic decay
with the characteristic $-\alpha/z'$ exponent. At longer times, however, the drift
term causes the exponent to decrease to $-\alpha$, resulting in a faster decay of
temporal correlations. 
The crossover time at which the transition between the two regimes occurs at
$\tau'_c \sim (B^{z'}/c_2)^{1/(1-z')}$.

We should note that the \emph{closed} 2D bosonic system in thermal equilibrium,
in the absence of drive and dissipation, reveals a slightly different
behaviour. In such a case the phase fluctuations obey the following equation: $
\partial_t^2\theta = D_x'\partial_x^2 \theta + D_y'\partial_y^2\theta,
$
where $D_x',D_y'$ are the squares of the $x$ and $y$-component,
respectively, of the speed of sound \cite{PhysRevX.5.011017,sieberer2015thesis}. In such a case,
 the two point correlation function shows an algebraic order with the
same exponent $\alpha$ for both space and time: $g^{(1)}(\mathbf{r},t)
\to \tilde{r}^{-\alpha},t^{-\alpha}$ respectively, with $\alpha>0$
\cite{szymanska2007mean}.  This is a consequence of
the linear dispersion of the gapless Bogoliubov excitation in
$k,\omega\to 0$ limit.

Including the compactness of the phase in the SA regime does not preclude
algebraic order and superfluidity. It leads to a well-known
BKT~\cite{PhysRevLett.111.088701} transition between a quasi-ordered and a
disordered phases mediated by the binding/unbinding of vortices. The system
falls into the XY universality class, which is the extension of the EW
universality class for compact variables.
We can estimate the phase boundary for algebraic order by considering a simple
argument presented in Ref.~\cite{PhysRevLett.111.088701,PhysRevX.5.011017}: we
assume that vortices only become relevant at scales where $g$ has flowed to
nearly $0$, and hence we can use the RG flow equations~\eqref{eq:rg_flow_eq}
even though they do not include vortices.\footnote{\label{SA_vortices}We note,
  however, that even in the SA regime the non-linearities might induce screening
  of the vortex interaction with a screening length that is shorter than the
  scale at which $g \approx 0$. It is an interesting question for future
  research whether this affects the BKT transition.} In this scenario, the BKT
transition is estimated to occur at $\kappa(\infty) = \pi$, where
$\kappa(\infty)$ is the renormalized scaled noise~\eqref{eq:ren_scaled_noise} in
the limit $l \to \infty$.  This condition defines the phase boundary
$\kappa_0 = \kappa_*$ between ordered and disordered phases which reads:
\begin{equation}
  \kappa_* = -\frac{4\pi\Gamma_0}{\left( 1 - \Gamma_0 \right)^2},
  \label{eq:bkt_transition}
\end{equation} 
where we have used the expression of $\kappa(\infty)$ as a function of the bare
parameters $\kappa_0$ and $\Gamma_0$~\cite{PhysRevLett.111.088701}. Thus, when
$\kappa_0 < \kappa_*$, the system shows algebraic order, whereas if
$\kappa_0 > \kappa_*$ the algebraic order is destroyed by vortices resulting in
exponential decay of correlations.

According to the above discussion, while incoherently pumped (and thus at best
weakly anisotropic) 2D polaritons (or other photonic) systems are always
disordered in the thermodynamic limit of infinite system size, and algebraic
order or superfluidity can only be a finite size effect, the parametrically
pumped polaritons are fundamentally different. The pumping process, which can be
chosen at any wave-vector, can result in a high level of effective anisotropy,
which allows us to enter the SA regime, governed by the $XY$ equilibrium fixed
point, thus ensuring algebraic order up to infinite distances in the
thermodynamic limit. Such a high level of anisotropy would not be achievable by
a crystal growth engineering aimed at creating different effective masses in
perpendicular directions. Moreover, as we show below, the anisotropy can be
changed simply by tuning experimental parameters such as the pump power or the
detuning between the excitons and photons, allowing us to easily in one
experiment move between different regimes.

%---- Figure  ----------------

\begin{figure}
  \includegraphics[width=8.3cm]{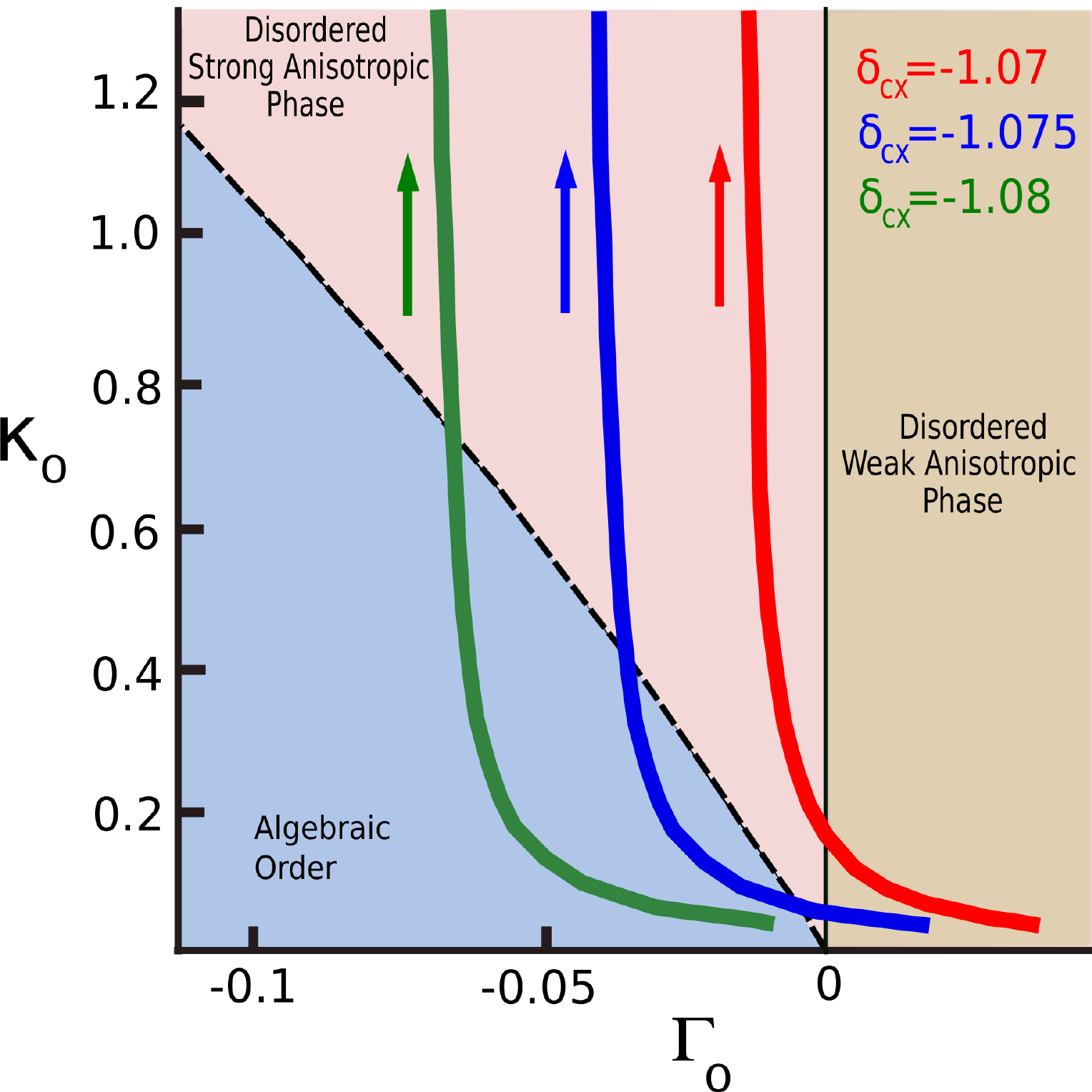}
  \caption{\textbf{Crossover between non-equilibrium and equilibrium-like
      universal regimes}.  Stable three-mode OPO configurations in the
    $\Gamma_0-\kappa_0$ space for three different detunings
    $\delta_{\mathit{CX}}=-1.07$ (green), $-1.075$ (blue), $-1.08$ (red),
    $k_p=1.4$ and $k_s=0.1$ (region A in Fig.~\ref{fig:mean_field}). The arrows
    indicate the direction of increasing the external drive strength. The dashed
    line shows the BKT phase boundary (see Eq.~\eqref{eq:bkt_transition})
    between algebraically ordered and the disordered phases.  By increasing the
    external pump power, for $\delta_{\mathit{CX}}=-1.07$ we cross from the
    non-equilibrium (WA) to the equilibrium-like (SA) disordered regimes, for
    $\delta_{\mathit{\mathit{CX}}}=-1.075$ the systems shows \emph{reentrance},
    it starts in the WA, enters the SA algebraically ordered
    regime and finally goes back to the disordered but now SA regime. For
    $\delta_{\mathit{CX}}=-1.08$ we are in the SA equilibrium-like regime for
    all pump powers, and the system undergoes a BKT transition from an
    algebraically ordered to a disordered phase by increasing $F_p$.}
\label{fig:anisotropic_regime_infinity_system}
\end{figure}

%---- End of Figure ----------------

%%%%%%%%%%%%%%%%%%%%%%%%%%%%%%%%%%%%%%%%%%%%%%%%%%%%%%%%%%%
%%%%%%%%%%%%%%%%%%%%%%%%%%%%%%%%%%%%%%%%%%%%%%%%%%%%%%%%%%%
\section{Exploring scaling regimes of OPO polaritons}
%%%%%%%%%%%%%%%%%%%%%%%%%%%%%%%%%%%%%%%%%%%%%%%%%%%%%%%%%%%
%%%%%%%%%%%%%%%%%%%%%%%%%%%%%%%%%%%%%%%%%%%%%%%%%%%%%%%%%%%
\label{sec:results}

In this section we show that the OPO-polariton system can be driven from the
non-equilibrium WA to an equilibrium-like SA regime by simply tuning the
strength of the external pump power and the detuning between the cavity photons
and excitons. We then consider implications for finite size systems. In general,
polaritons in the OPO regime (or in the incoherently pumped scenario above
condensation threshold) are characterised by a high degree of coherence. This is
because the system size in experiments, due to intrinsic disorder in the samples
limiting the spatial extent of useful regions, is relativity small in comparison
to the relevant length scales of the decay of correlations, especially well
above threshold where most experiments operate.
Indeed, non-decaying spatial coherence, characteristic of BEC in 3D, was seen in most cases~\cite{kasprzak2006bose,spano2013build}, and even observation of the algebraic decay appeared challenging~\cite{roumpos2012power,nitsche2014algebraic,caputo2016topological}. In
order to minimise the influence of the finite size, which masks the underlying
physics, we need to focus on samples and regimes, in which 
we have appreciable decay of coherence for the considered system size.
In general, this would correspond to what we call \textit{bad} samples, where the
influence of dissipation is substantial but not strong enough to completely wash
out any collective effects. In our opinion, the most promising microcavities are
those used in early days of work on polariton condensation, where collective
effects were already seen but the polariton lifetime and the Rabi splitting were
quite small by current standards.
Thus, we first focus on what we call a \textit{bad} inorganic
microcavity~\cite{sanvitto2010persistent,dagvadorj2015nonequilibrium}, and in
Sec.~\ref{sec:experiments} we compare this with better quality
samples, characterised by longer lifetimes and larger Rabi splitting,
used by most groups today, as well as with organic microcavities.

The \textit{bad} microcavity is characterized by the following set of parameters:
$\hbar\Omega_R=4.4 \, \mathrm{meV},\; \hbar\gamma_j=0.1 \, \mathrm{meV}$
(corresponding to a lifetime of $6.6 \, \mathrm{ps}$);
$m_C=2.5 \cdot 10^{-5} m_0, \; g_X=2 \mu\textrm{eV} \mu \mathrm{m}^{-2}$.
We choose the pump wave vector close to the inflection point of the
lower polariton dispersion, $k_p=1.4$, which corresponds to $1.61
  \; \mu \mathrm{m}^{-1}$ in dimensional units for the \textit{bad} cavity, and
$\omega_p=\omega_{\mathrm{LP}}(k_p)$, as shown in
Fig.~\ref{fig:mean_field}. At the mean-field level, the system
exhibits upper and lower thresholds for the OPO transition at pump
powers indicated by $F^{\textrm{lo}}_p$ and $F^{\textrm{up}}_p$
respectively. Solving numerically exactly the analogue of
Eq.~\eqref{eq:SCGPE} for the exciton-photon model and the same set of
parameters with zero-detuning shows that the system undergoes a BKT-type phase transition
at a pump power $F^{\textrm{BKT,lo}}_p\approx 1.014F^{\textrm{lo}}_p$
and $F^{\textrm{BKT,up}}_p\approx 0.999F^{\textrm{up}}_p$
\cite{dagvadorj2015nonequilibrium}. The value of $k_s$ is not
determined by the three-modes ansatz~\eqref{eq:SCGPE}
\cite{dunnett2017polariton}.
However, the stability analysis shown in Fig.~\ref{fig:mean_field} suggests a value of
$k_s \approx 0.11 \, \mu \mathrm{m}^{-1}$ for intermediate and high values of
$F_p$, whereas the system chooses bigger values of $k_s$ when approaching the
lower threshold, i.e.,
$k_s\in[0.11 \, \mu \mathrm{m}^{-1},0.7 \, \mu \mathrm{m}^{-1}]$ (region C in
Fig.~\ref{fig:mean_field}).

%%%%%%%%%%%%%%%%%%%%%%%%%%%%%%%%%%%%%%%%%%%%%%%%%%%%%%%%%%%
\subsection{Infinite system: crossover between weakly and strongly anisotropic
  regimes}
%%%%%%%%%%%%%%%%%%%%%%%%%%%%%%%%%%%%%%%%%%%%%%%%%%%%%%%%%%%
\label{sec:infinite_crossover}

%---- Figure non-zero detuning ----------------

\begin{figure}
  \includegraphics[width=0.238\textwidth]{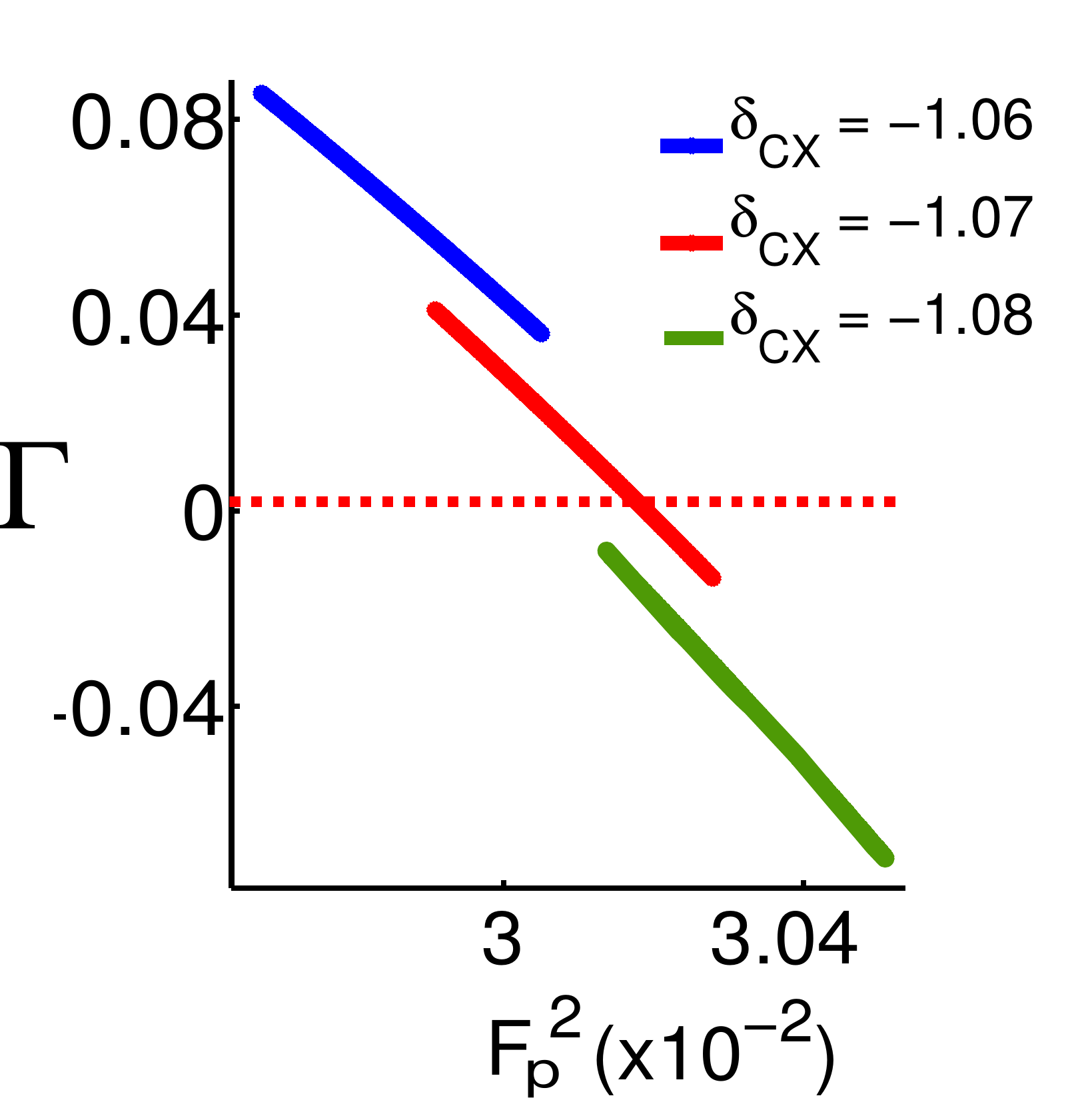}
  \includegraphics[width=0.238\textwidth]{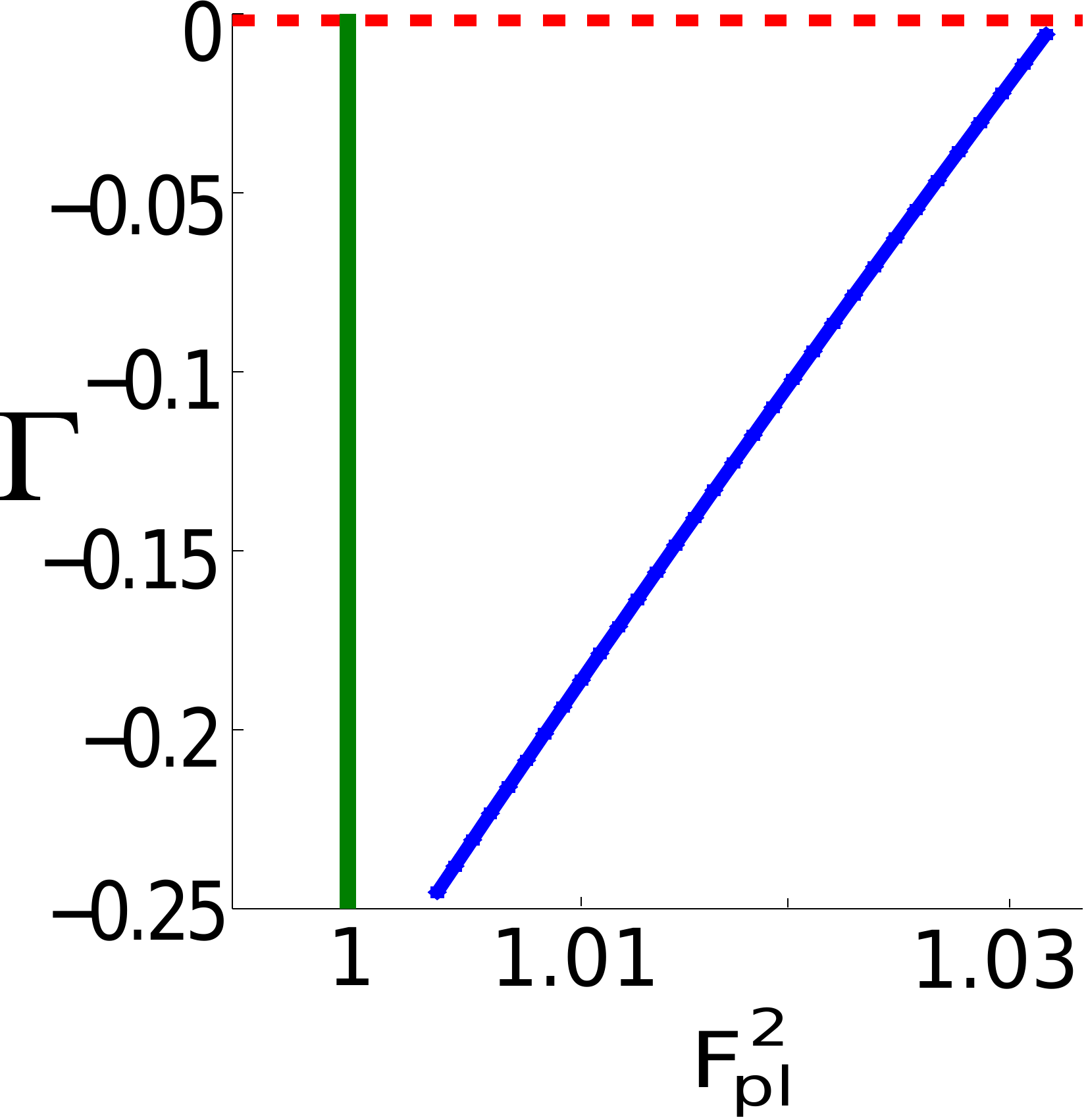}
  \caption {\textbf{Changing the effective anisotropy by tuning the driving
      strength}.  \emph{Left panel:} Anisotropy parameter $\Gamma$ as a function
    of pump power $F_p^2$ (in the same range as region A in
    Fig.~\ref{fig:mean_field}) for different detunings
    $\delta_{\mathit{CX}} = -1.06$ (blue), $-1.07$ (red), $-1.08$ (green) in the
    stable three-mode OPO regime.  Parameters are as in
    Fig.~\ref{fig:anisotropic_regime_infinity_system}.  For
    $\delta_{\mathit{CX}} = -1.075$ the anisotropy parameter crosses zero
    (dotted red line). Note that the extent of region A, i.e., the range of
      pump powers for which the three-mode OPO ansatz is stable, depends on the
      value of the detuning.  \emph{Right panel:} Anisotropy parameter $\Gamma$
    as a function of normalised pump power, where
    $F_{\mathit{pl}}\equiv F_p/F^{\textrm{lo}}_p$, at zero detuning but close to
    the \emph{lower} OPO threshold, indicated by the green vertical line.  The
    pump and signal have large momenta $k_p \approx 1.84$ and $k_s =1.0$. The
    dotted red line marks the $\Gamma=0$ line. This case corresponds to region C
    in Fig.~\ref{fig:mean_field}.}
\label{fig:isotropic_phase_non_zero_detuning}
\end{figure}

%---- End of Figure ----------------

We first focus on the region with small $k_s$ (A and B in
Fig.~\ref{fig:mean_field}). The analysis of the anisotropy parameter $\Gamma$
for the \textit{bad} cavity shows that the system falls into the WA regime, i.e.,
$\Gamma>0$, at all pump powers when the detuning between the photons and the
excitons fulfils $-1.07 < \delta_{\mathit{CX}}$. The expected scaling behaviour
of correlations is discussed in Sec.~\ref{sec:weakly-anis-regime}. In
particular, algebraic decay of correlations is ruled out in this regime.
However, this picture changes drastically for lower values of the detuning.
When $\delta_{\mathit{CX}}\leq-1.07$, we enter the SA regime, which is expected
to have long-range properties which are qualitatively similar to an equilibrium
system (see Sec.~\ref{sec:strongly-anis-regime}). For example, algebraic order
can be destroyed by the proliferation of vortices as in the equilibrium BKT
transition when the level of the effective noise is large, which is the case when
the signal density is low. In Fig.~\ref{fig:anisotropic_regime_infinity_system}
we show three characteristic cases of different detuning superimposed on the
phase diagram generated by the relation~\eqref{eq:bkt_transition} in the
$\Gamma_0-\kappa_0$ phase-space: i) For $\delta_{\mathit{CX}} = -1.07$ the
system is always disordered. It crosses from the disordered non-equilibrium (WA)
to the disordered equilibrium-like (SA) regime since $\kappa_0>\kappa_*$ for all
values of $F_p$. ii) $\delta_{\mathit{CX}} = -1.075$ is the most interesting
case. The system shows reentrant behaviour \cite{PhysRevX.5.011017}: by
increasing the pump power we move from the disordered WA to the disordered SA
regime, then by increasing the pump power further we reach the BKT phase
transition to the algebraically ordered phase, followed by a second BKT
transition close to the OPO upper threshold back to the SA disordered
phase. iii) For $\delta_{\mathit{CX}} = -1.08$ the stable three-mode solutions
lie in the SA regime for all pump powers. For smaller values of $F_p$
the system is in the algebraically ordered phase, and it undergoes a BKT phase
transition to a disordered phase by increasing $F_p$. As can be seen in the
left panel of Fig.~\ref{fig:isotropic_phase_non_zero_detuning}, all these
three different cases show a nearly linear dependence of $\Gamma$ on the
intensity $F^2_p$ of the external drive.

Increasing detuning in the region close to the upper threshold (region
A in Fig.~\ref{fig:mean_field}) is one way to introduce a sufficient
level of anisotropy to cross to the equilibrium-like phase.  However,
there is also another source of anisotropy: the system can be driven
to the SA regime by increasing the pump momentum, $k_p$, leading to an
increase of the signal momentum, $k_s$, by tuning the pump power
close to the lower OPO threshold (region C in
Fig.~\ref{fig:mean_field}).  For example, for the \textit{bad} cavity
parameters with $\delta_{\mathit{CX}}=0$ and $k_p=2.11 \, \mu
\mathrm{m}^{-1}$ (1.84 in dimensionless units) we enter the SA regime
for $k_s\approx0.7 \, \mu \mathrm{m}^{-1}$, as can be seen in the
right panel of Fig.~\ref{fig:isotropic_phase_non_zero_detuning}.

%%%%%%%%%%%%%%%%%%%%%%%%%%%%%%%%%%%%%%%%%%%%%%%%%%%%%%%%%%%%%%%%%%%
\subsection{Finite system: the length scales of the weakly anisotropic regime}
\label{sec:finite-syst-length}
%%%%%%%%%%%%%%%%%%%%%%%%%%%%%%%%%%%%%%%%%%%%%%%%%%%%%%%%%%%%%%%%%%%

%----- Figure----------------

\begin{figure*}
  \includegraphics[width=5.8cm]{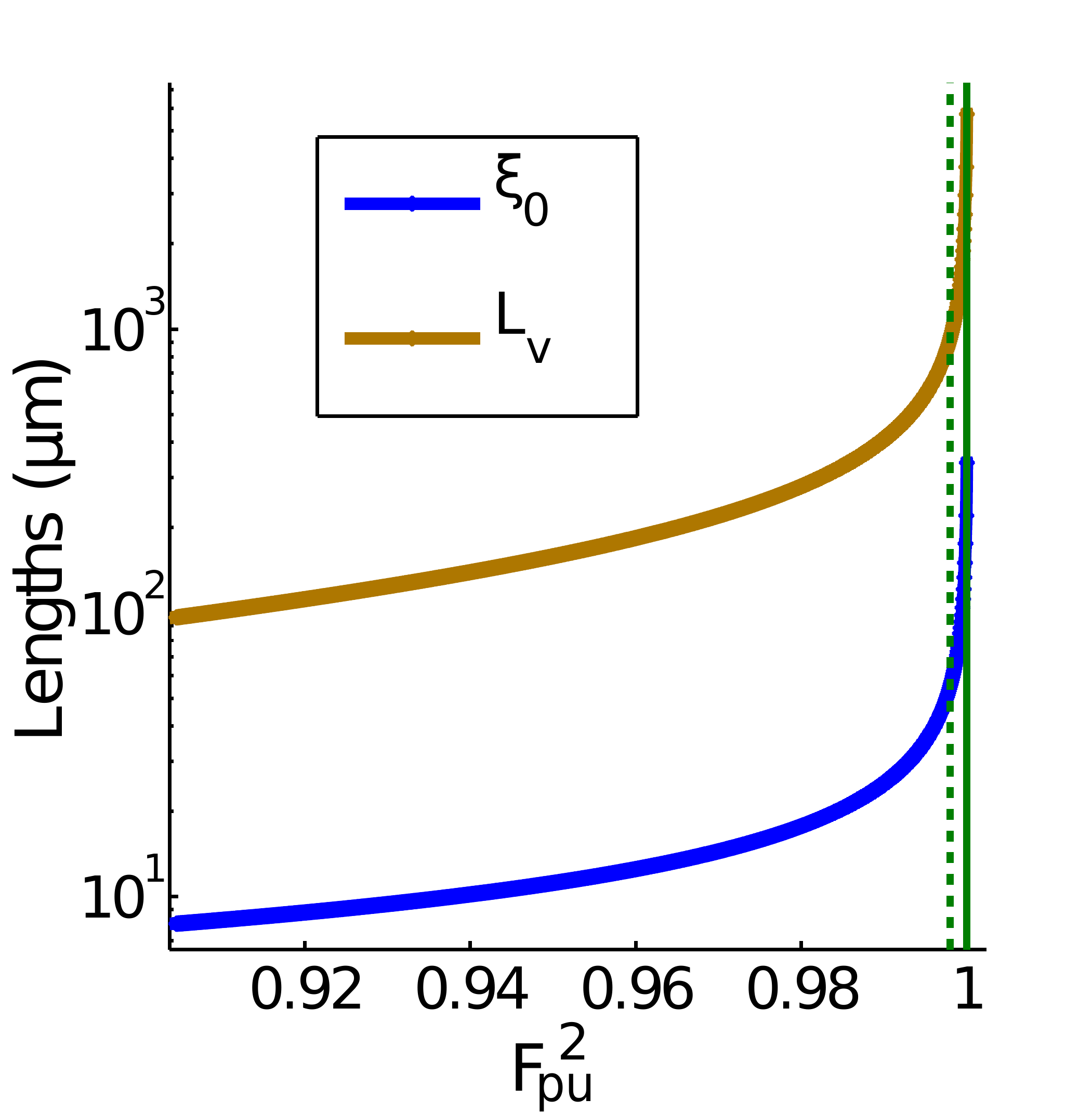}
  \includegraphics[width=5.8cm]{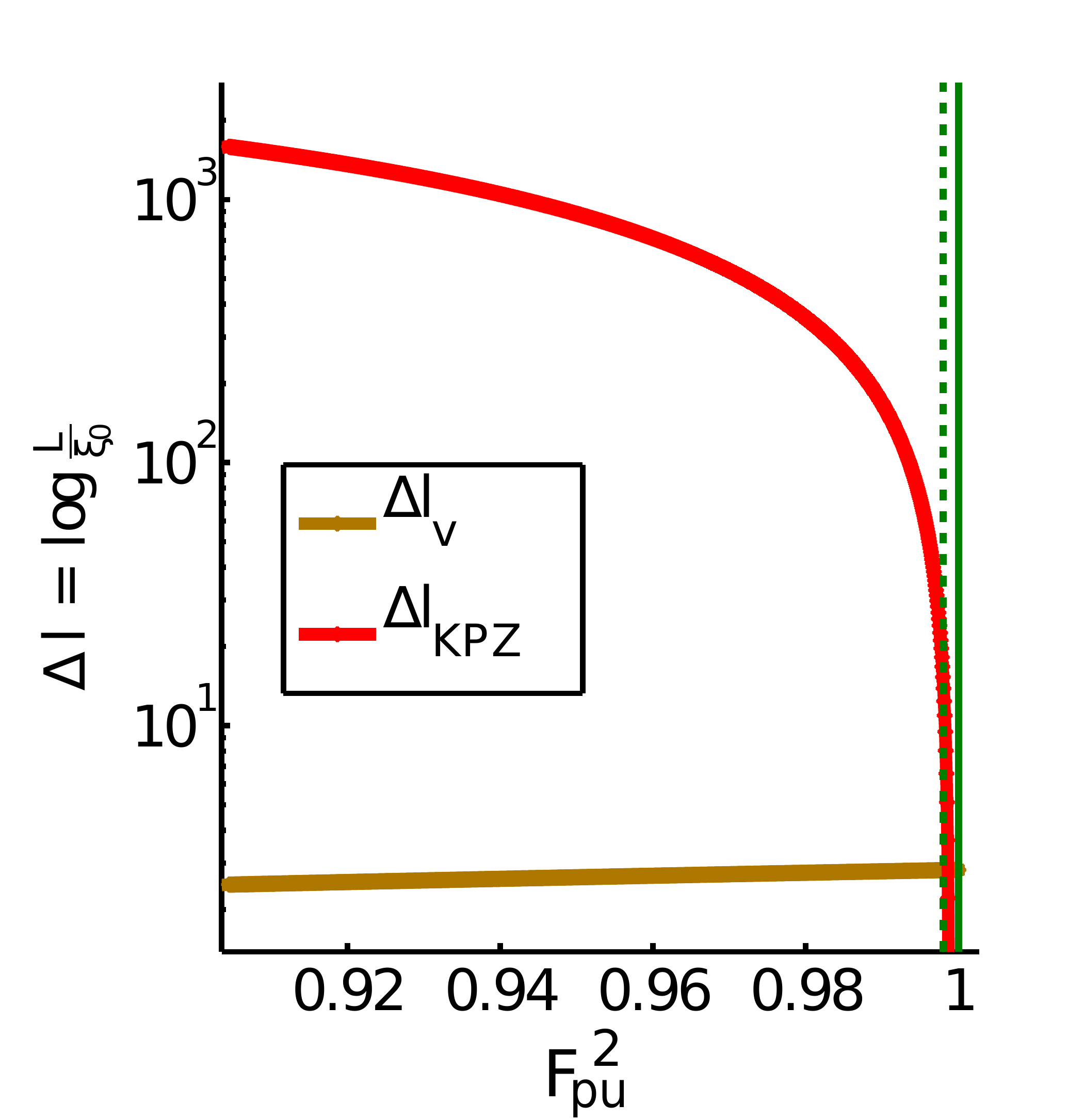}
  \includegraphics[width=5.8cm]{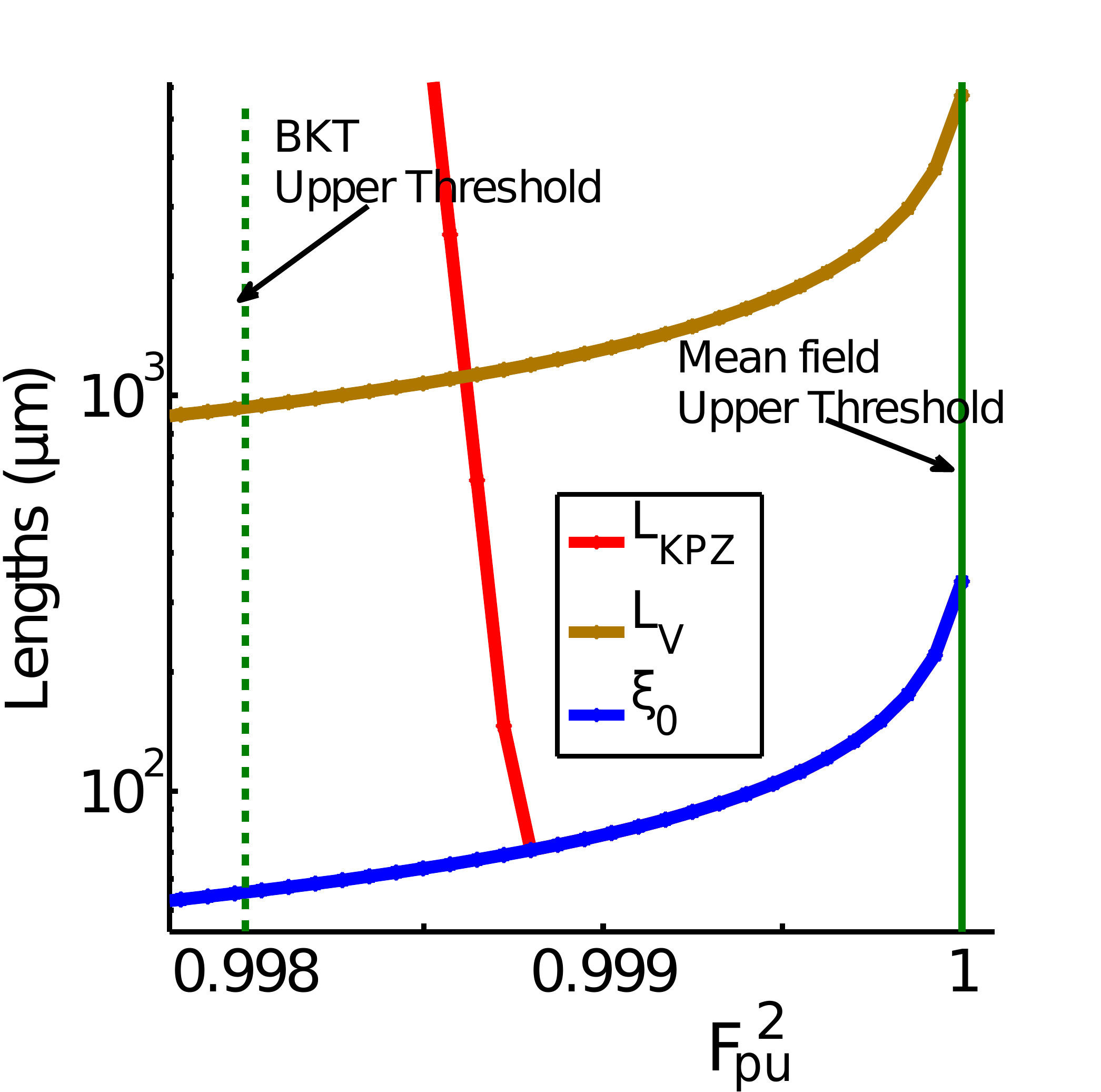}
  \caption  { {\bf Characteristic length scales in the WA regime.}
    \emph{Left and right panels}: Healing length $\xi_0$ (blue), length scale
    for the vortex-dominated disordered phase $L_v$ (brown), and the KPZ length
    scale $L_{\mathrm{KPZ}}$ (red) as a function of the pump strength
    (normalised to the upper threshold,
    $F_{\mathit{pu}}\equiv F_p/F^{\textrm{up}}_p$). Vertical dotted (green) line
    indicates the BKT transition where V-AV pairs proliferate (taken from
    Ref.~\cite{dagvadorj2015nonequilibrium}) and the vertical solid (green) line
    the mean-field OPO threshold. Parameters as those discussed in the text for
    the \textit{bad} microcavity at zero detuning $\delta_{\mathit{CX}}=0$, $k_p=1.4$ and
    $k_s=0.1$. The two different panels correspond to two different ranges in
    pump strength. \emph{Central panel}: logarithmic RG scale for the
    vortex-dominated phase ($\Delta l_v$) and for the KPZ-phase
    ($\Delta l_{\mathrm{KPZ}}$) as a function of pump strength
    normalised to the pump at the upper threshold (vertical solid green
    line). Vertical dotted green line indicates the BKT transition.
    Note that $L_v$ is expected to be strongly renormalized close to the threshold as explained in Appendix \ref{sec:appendix_vortex}.
}
\label{fig:lenghts_galaa_parameters}
\end{figure*}

% ----- End Figure----------------

In this subsection, based on Eqs.~\eqref{eq:length_KPZ}
and~\eqref{eq:vortex_scale}, we estimate the relevant length scales of the WA
regime to examine which phases can be seen in current semiconductor
microcavities, and whether finite-size effects would hamper the underlying
universal physics. We address, in particular, whether the non-equilibrium
ordered KPZ phase can ever be seen in semiconductor microcavities. We focus here
on our \textit{bad} cavity parameters as the most promising to explore various phases.

We first explore the WA regime close to the upper threshold (region A
in Fig.~\ref{fig:mean_field}). The parameters of the aKPZ equation for
this regime are shown in Fig.~\ref{fig:kpz_parameters_galaa} in
Appendix~\ref{sec:appendix_kpz}.  Most of the parameters are
approximately constant as a function of the external pump strength,
$F_p$. The drift term is non-zero only in the direction of the pump
wave vector $\mathbf{k}_p$, which we have chosen to be along the
$x$-axis. The dimensionless non-linearity $g$ and the noise strength
$\Delta$ asymptote to high values, which leads to a small value of
$L_{\mathrm{KPZ}}$ (see Eq.~\eqref{eq:length_KPZ}), only very close to
the upper threshold. As can be seen in
Fig.~\ref{fig:lenghts_galaa_parameters}, even for the most promising
\textit{bad} cavity parameters $L_{\mathrm{KPZ}}$ is astronomically large
at any reasonable distance from the upper threshold. It goes down to
$100 \, \mu \mathrm{m}$, currently the upper bound for any
experiments, only at around 0.999 of the upper threshold (see right
panel of Fig.~\ref{fig:lenghts_galaa_parameters}). However, this point
is already above the BKT transition (green dashed line in
Fig.~\ref{fig:lenghts_galaa_parameters}), where proliferating
vortex-antivortex pairs destroy the KPZ scaling. Additionally, $L_v
\ll L_{\mathrm{KPZ}}$ for all values of $F_p$ apart from those close
to the upper threshold already beyond the BKT transition. (Also, note
that as explained in Sec.~\ref{sec:weakly-anis-regime} close to the
upper threshold where fluctuations are strong the scale at which
vortices unbind should be strongly renormalized and smaller than $L_v$
in Eq.~\eqref{eq:vortex_scale}.) Thus, we conclude that in this regime
(region A in Fig.~\ref{fig:mean_field}) KPZ scaling would either be
overshadowed by the algebraic order at scales below $L_v$ due to the
astronomically large length scales required, or destroyed by the BKT
vortices resulting in an equilibrium-like behaviour in a finite
system. The question remains: is there then no hope for the KPZ phase
in microcavities in two dimensions and we are only left with equilibrium-like behaviour?
We address this in the next section.

We should also comment that $L_v$ drops down to less then $100 \, \mu
\mathrm{m}$ for our \textit{bad} cavity for some pump powers away from the
BKT threshold --- a scale which is quite realistic.
This suggests that in such a case free vortices (not of BKT type)
should destroy the algebraic order beyond this scale
\cite{wachtel2016electrodynamic} (see Appendix
\ref{sec:appendix_vortex}). However, exact simulations of stochastic
dynamics for systems as large as $1000 \, \mu \mathrm{m}$
\cite{dagvadorj2015nonequilibrium} do not show any signs of this
phase, even at very long times where a steady-state has clearly been
reached, suggesting suppressed activation by e.g. an extremely small
vortex mobility (Ref. \cite{wachtel2016electrodynamic} assumed instead
a vortex mobility of the order of other scales in the problem), or
that attractive interactions between vortex and antivortex at small
distances may in reality prevent the free vortices from becoming
relevant. It may also be that the rough formula for $L_v$
underestimates the real value.

%%%%%%%%%%%%%%%%%%%%%%%%%%%%%%%%%%%%%%%%%%%%%%%%%%%%%%%%%%%%%%%%%%%
\subsection{Finite system: searching for the KPZ phase}
\label{sec:finite-syst-search}
%%%%%%%%%%%%%%%%%%%%%%%%%%%%%%%%%%%%%%%%%%%%%%%%%%%%%%%%%%%%%%%%%%%

%---- Figure ----------------

\begin{figure}
  \includegraphics[width=0.22\textwidth]{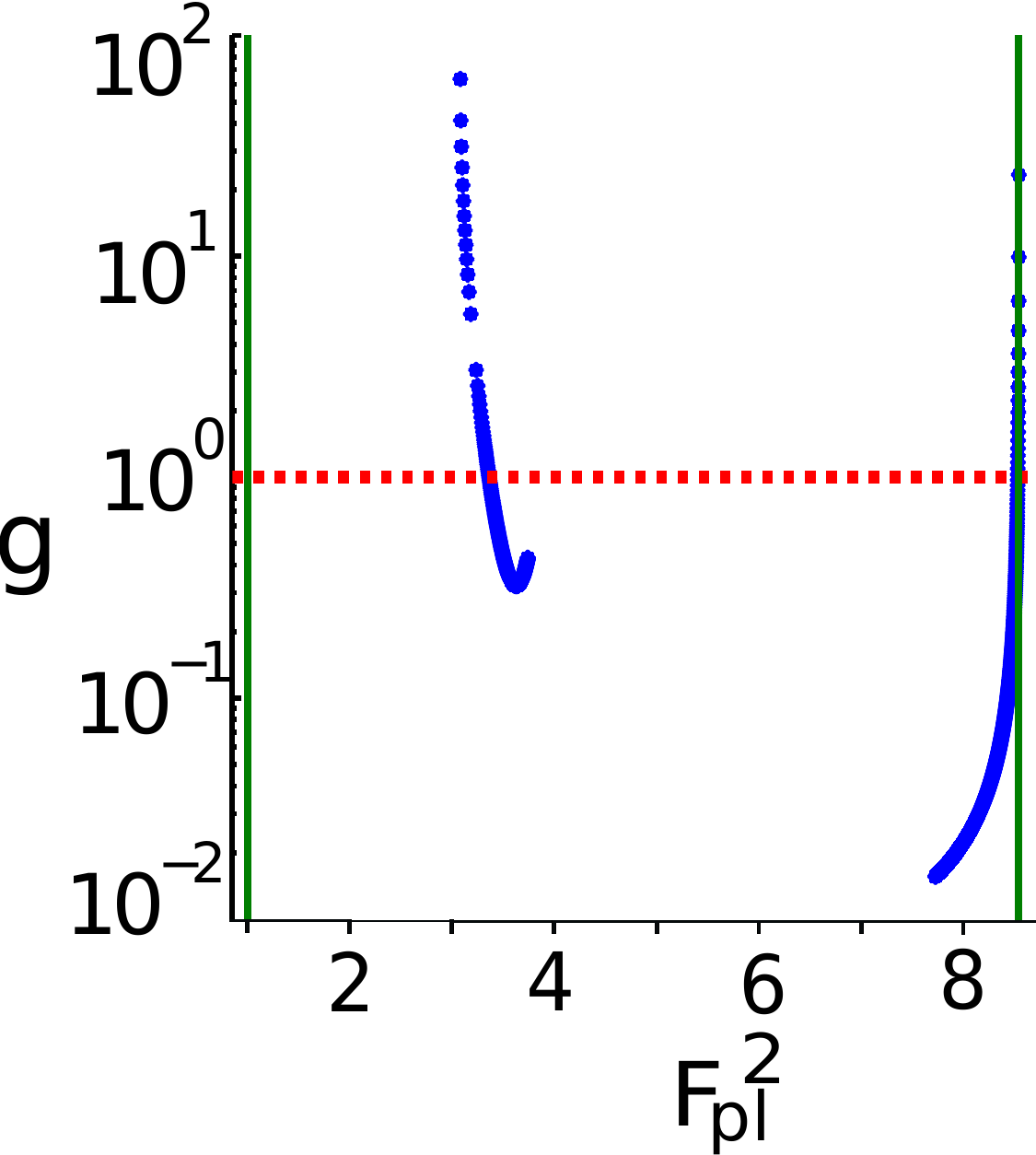}
\includegraphics[width=0.235\textwidth]{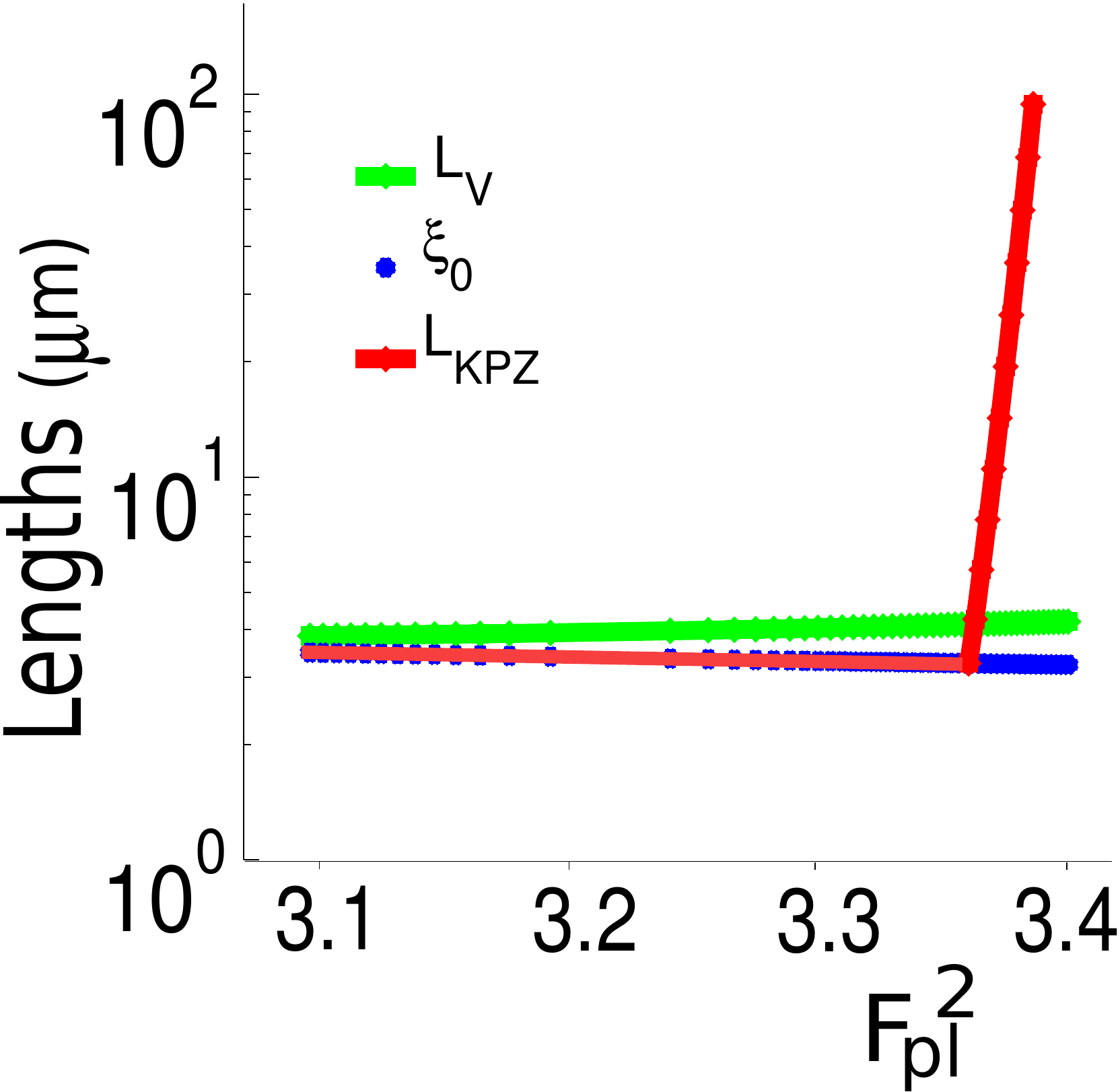}
  \caption { \textbf{Characteristic length scales in the WA regime at
      intermediate pump powers.}  \emph{Left panel}: Non-linear parameter $g$ as
    a function of the pump strength normalised to the pump at the lower
    threshold for stable three-mode OPO solution with zero detuning. The
    vertical green lines indicate the lower and upper mean field thresholds, and
    the red dotted line marks the $g=1$ border. These are the stable solutions
    at intermediate values of $F_p$ (region B in Fig.~\ref{fig:mean_field}).
    \emph{Right panel}: Characteristic length scales at the intermediate $F_p$
    shown in the left panel.  Healing length $\xi_0$ (blue), length scale for
    vortex dominated disordered phase $L_v$ (green) and the KPZ length scale
    $L_{\mathrm{KPZ}}$ (red) as a function of pump strength normalised to the
    pump at the upper threshold.  Note that to the left of the nearly
      vertical increase of $L_{\mathrm{KPZ}}$ its value is indistinguishable
      from the healing length. }
\label{fig:lengths_intermediate_fp_galaa}
\end{figure}

%---- End of Figure ----------------

Unlike the incoherently pumped case, our OPO system offers more possibilities for parameter tuning.
Interestingly, as we can see in Fig. \ref{fig:lengths_intermediate_fp_galaa},
the system shows a regime at intermediate pump powers,
$3.1<(F_p/F^{\textrm{lo}}_p)^2<3.5$, where $g$ becomes large and so
$L_{\mathrm{KPZ}}$ is small (region B in Fig. \ref{fig:mean_field}). In fact, in
some parts of this range $g>1$, meaning that the KPZ phase is expected at all length
scales beyond the healing length.
Such a regime does not exist for incoherently
driven microcavities. Its presence in the OPO configuration is due to the
non-monotonic behaviour of KPZ parameters as a function of pump power,
associated with underlying instabilities towards more complex spatial patterns
in the system such as the satellites formation or ring OPOs \cite{dunnett2017polariton}.
However, we also find that $L_v\approx\xi_0$ (see right panel
Fig. \ref{fig:lengths_intermediate_fp_galaa}) and therefore in principle the
vortex phase could destroy the KPZ physics. 
But, due to the suppressed activation observed in numerical studies,
the free vortices may never appear. Testing the possibility of the KPZ
phase in the middle of the OPO region using the exact stochastic
dynamics, and hopefully experiments, would give the final word on
this. Note, that the only experiment measuring spatial
  coherence in the OPO configuration focuses on a different regime of
  powers \cite{spano2013build}. Our estimates show that using pump powers a few times
  the OPO threshold in good quality samples as far as spatial
  disorder is concerned, but with relatively short polariton lifetime, is
  the most promising regime to observe signatures of the KPZ physics.

%%%%%%%%%%%%%%%%%%%%%%%%%%%%%%%%%%%%%%%%%%%%%%%%%%%%%%%%%%%%%%%%
\subsection{Finite system: crossover between weakly and strongly
  anisotropic regimes}
\label{sec:finite-syst-cross}
%%%%%%%%%%%%%%%%%%%%%%%%%%%%%%%%%%%%%%%%%%%%%%%%%%%%%%%%%%%%%%%%

%---- Figure ----------------

\begin{figure}
\begin{center}
  \includegraphics[width=0.5\textwidth]{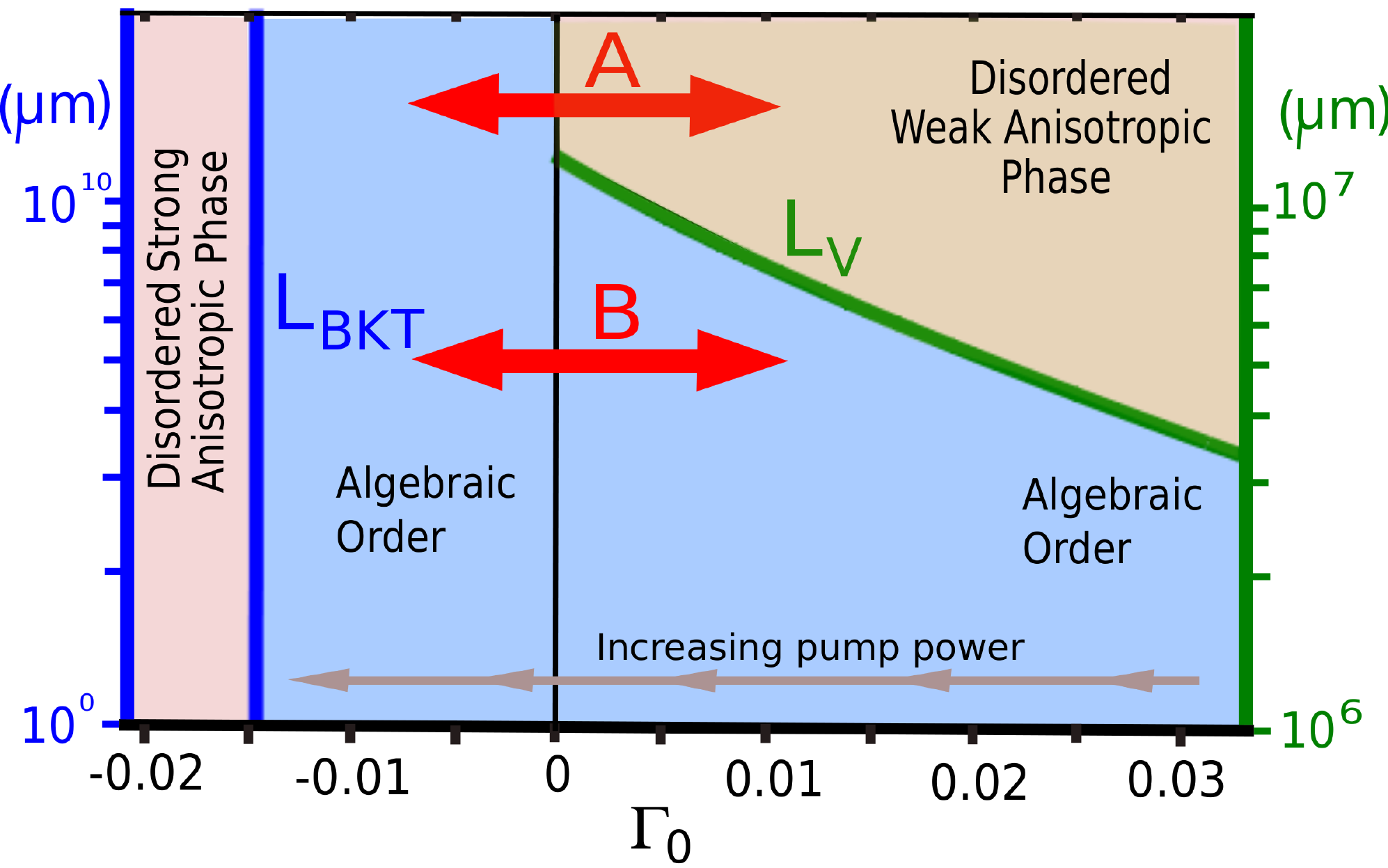}
  \caption { \textbf{Crossover between SA (negative $\Gamma_0$) and WA (positive
      $\Gamma_0$) regimes at finite length scales}.  Length scale for the vortex
    dominated phase in the WA regime $L_v$ (green line) and the disordered phase
    in the SA $L_{\textrm{BKT}}$ (blue line) as a function of the anisotropy
    parameter, tunable by the drive, for the polariton system with
    $\delta_{\mathit{CX}} = -1.07$. The double red arrow A indicates a
    transition from the disordered WA to an algebraically ordered SA phase by
    increasing the external pump power in a system of size $L$ such that
    $L_v<L<L_{\textrm{BKT}}$. The double red arrow B indicates a transition from
    the ordered WA regime to the ordered SA regime by increasing the external
    pump power in a system of size $L$, where $L<L_v<L_{\textrm{BKT}}$.}
\label{fig:strong_anisotropy_finite_lengths}
\end{center}
\end{figure}

%---- End of Figure ----------------

In Sec.~\ref{sec:infinite_crossover} we have shown that the infinite polariton
system can be tuned between two different universality classes (non-equilibrium
KPZ and equilibrium EW), with completely different large-scale behaviours, by
changing the exciton-photon detuning, and the properties of the drive such as
the pump power, $F_p$, and wave-vector $k_p$ --- easily realisable in current
experiments. Here, we consider how this crossover is affected by the finite
size.

We consider the case with detuning $\delta_{\mathit{CX}} = -1.07$
close to the upper OPO threshold (region A in
Fig. \ref{fig:mean_field}). As we discussed, for this detuning the
system can move from the non-equilibrium (WA) to an
equilibrium-like (SA) regime by increasing the external pump
power. If the system is infinite, for these parameters, it is in the
disordered phase, characterised by the exponential decay of
correlations, in both regimes.
 However, the system may show algebraic order up to certain
  length scale $L_{\textrm{BKT}}$ in the SA
    regime or $L_v,L_{\textrm{KPZ}}$ in the WA
    regime.  In the first case, this would require
  $\kappa_*<\kappa(l)<\pi$ (see \eqref{eq:bkt_transition}), and
  $L_{\textrm{BKT}}\equiv\xi_0e^{l_{\textrm{BKT}}}$ is obtained by
  considering the 'BKT-transition criterion':
  $\kappa(l_{\textrm{BKT}})\sim\pi$.
  We calculate $L_{\textrm{BKT}}$ following
  Ref. \cite{PhysRevLett.111.088701} and the RG flow-equations
  \eqref{eq:rg_flow_eq}, and the results are displayed in
  Fig. \ref{fig:strong_anisotropy_finite_lengths}. We obtain that
  $L_{\textrm{BKT}}$ takes reasonable physical values for
  $\Gamma\approx -0.0145$.  When $\Gamma> -0.0145$,
  $L_{\textrm{BKT}}\to \infty$ which indicates that the algebraic
  order appears at all realistic physical length scales; whereas when
  $\Gamma < -0.0145$, $L_{\textrm{BKT}}\sim \xi_0$, and so the system
  is in a disordered phase at all length scales beyond the healing
  length.  

Considering this, we find that there are two interesting scenarios at
intermediate length scales when driving the system from the WA to the
SA regime, as indicated by double arrows $A$ and $B$ in
Fig. \ref{fig:strong_anisotropy_finite_lengths}. The first case, arrow $A$,
shows a transition between the disordered phase in the non-equilibrium WA
regime to the algebraically ordered phase in the equilibrium-like SA
regime by increasing the pump power and, consequently, crossing $\Gamma=0$. This
phenomenon appears for $L$ such that $L_v<L<L_{\textrm{BKT}}$.  In the second scenario (arrow
$B$ in Fig. \ref{fig:strong_anisotropy_finite_lengths}), the system can be
driven from the WA to the SA regimes without changing the phase,
i.e. maintaining the algebraic order in both cases. This situation occurs for
$L$ fulfilling $L<L_v,L_{\textrm{BKT}}$.  Scenario $A$ requires length scales of the order of
meters for our \textit{bad} cavity parameters. Thus scenario $B$ is more likely in
current experiments. Realising scenario A would require increasing the
dissipation in a controlled way so that not all the collective effects are
washed out and the OPO survives.

\begin{figure}
  \includegraphics[width=0.24\textwidth]{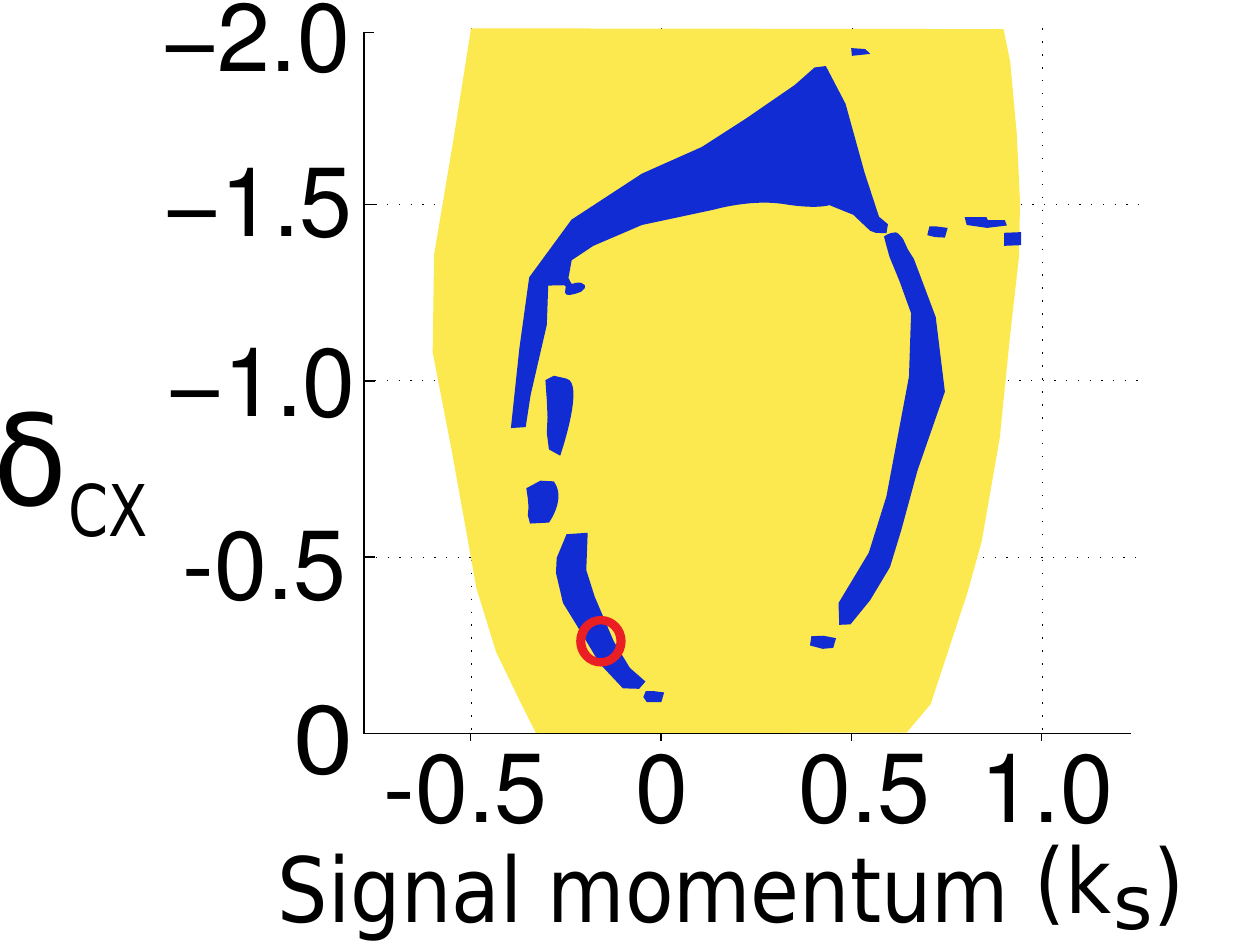}
  \includegraphics[width=0.22\textwidth]{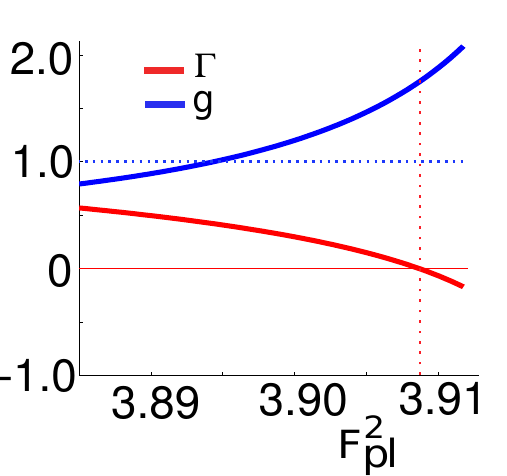}
  \caption { \textbf{Crossover between the KPZ scaling and the SA (negative
      $\Gamma_0$) regime with algebraic order.}  \emph{Left panel:} Regions of
    detuning between the cavity photons and the excitons $\delta_{\mathit{CX}}$
    and signal momentum $k_s$ (dark blue) for which the system can exhibit a
    crossover between the KPZ phase and algebraically ordered phase by tuning
    the pump strength $F_p$. Yellow region marks stable KPZ-solutions.
    \emph{Right panel:} $g$ and $\Gamma$ as a function of the normalized pump
    strength $F^2_{\rm{pl}}$ for $k_s=-0.15$ and $\delta_{\mathit{CX}}=-0.25$
    (small red circle in the left pannel). The crossover between the KPZ and the
    SA regimes occurs at $F^2_{\rm{pl}}\simeq 3.909$: $\Gamma$ changes sign
    (vertical red dotted line) while $g>1$ (dotted blue horizontal line).
    $k_p=1.4$ and we consider all values of $k_s$ for which there is a non-zero
    mean field solution and a stable aKPZ equation. \textbf{Note:} Stability
    analysis indicates that in the blue regions the three-mode ansatz is
    unstable. This is confirmed by numerical
    studies~\cite{dunnett2017polariton}, which show that the steady-state
    polariton field can develop several secondary modes in addition to the three
    main ones. However, the secondary modes are typically orders of magnitude
    weaker, and we expect them to give only minor quantitative corrections to
    the values of $g$ and $\Gamma$ we find here.
  }
\label{fig:transition_kpz_gamma_neg}
\end{figure}

Finally, we can ask whether a transition between WA ($\Gamma>0$)
  and SA ($\Gamma<0$) regimes is possible in conditions of strong KPZ
  non-linearity ($g>1$) where KPZ scaling would show at all
  lengthscales beyond the healing length. This would mean crossing a
  phase with stretched exponential decay of correlations to a phase
  with algebraic order as in equilibrium systems below the BKT
  transition. We find that such KPZ to algebraic-order crossover as a
  function of pump power is indeed possible at finite detuning (see
  Fig. \ref{fig:transition_kpz_gamma_neg} for the \emph{bad cavity} system).

%%%%%%%%%%%%%%%%%%%%%%%%%%%%%%%%%%%%%%%%%%%%%%%%%%%%%%%%%%%
\subsection{Different experimental systems}
%%%%%%%%%%%%%%%%%%%%%%%%%%%%%%%%%%%%%%%%%%%%%%%%%%%%%%%%%%%
\label{sec:experiments}

In the previous section we studied in detail the \textit{bad} cavity configuration
characterised by relatively large photon decay rate with a polariton lifetime of
$\tau \approx 6.6 \, \mathrm{ps}$, as the most promising for observation of
different phases.  However, current state-of-the-art inorganic microcavities are
characterised by much longer polariton lifetimes. Indeed, \emph{typical}
inorganic cavities have polariton lifetimes of $\tau\approx 30 \, \mathrm{ps}$
\cite{wertz2010spontaneous}, 
whereas the best cavities show polariton lifetimes
$\tau\approx 150 \, \mathrm{ps}$ \cite{steger2015slow,caputo2016topological}. We
refer to the latter as \emph{good} cavities. Most inorganic samples are
characterised by Rabi splittings $\Omega_R$ and exciton-exciton
interaction-strengths $g_X$ comparable to the ones used in the previous section
for the \textit{bad} cavity, i.e., $\hbar\Omega_R\approx 4.4 \, \mathrm{meV}$ and
$g_X\approx 0.002 \, \mathrm{meV}\mu m^{-2}$, respectively.  On the other hand,
\emph{organic} microcavities have extremely low photon lifetime, but high
Rabi splittings and relatively small exciton-exciton interaction strength.
Typical values for organic cavities are
$\tau\approx 5.5\cdot 10^{-2}\, \mathrm{ps}$,
$\hbar\Omega_R\approx 1000 \, \mathrm{meV}$ and
$g_X\approx 10^{-6} \, \mathrm{meV}\mu m^{-2}$~\cite{daskalakis2014nonlinear}.

In this section we present a comparison of the relevant length scales for these
different cavities, i.e., the \textit{bad, typical, good} and organic cavities
with zero detuning between the cavity-photons and the excitons. We first
focus on the OPO region at high pump powers (region A in
Fig.~\ref{fig:mean_field}).  The results are listed in
Table~\ref{table:setups}. The first four rows show: the length scale for the
vortex dominated phase ($L_v$), the KPZ-phase ($L_{\mathrm{KPZ}}$), the healing
length ($\xi_0$), and the corresponding normalized pump power
$F_{\mathit{np}}^2$ at a point where the $L_v$ is smallest within stable
three-mode OPO solutions (see left panel in
Fig.~\ref{fig:lenghts_galaa_parameters} as as example for the \textit{bad}
cavity).  The estimate for $L_v$ is done using $\xi_0 \approx \bar{\xi}_0$
(cf. the discussion around Eq. \eqref{eq:vortex_scale}, which may not be
realistic, so this information has to be taken with a grain of salt. In
addition, we observe an $L_v$ which takes reasonable physical values for the
\textit{bad} cavity and organic samples, whereas the KPZ scale
$L_{\mathrm{KPZ}}$ is unreachable.

As we mentioned in the last section, close to the upper threshold, the KPZ
length scale drops down significantly, reaching $\xi_0$ and $L_v$ (See right
panel in Fig.~\ref{fig:lenghts_galaa_parameters} as an example for the
\textit{bad} cavity. Also, note that $L_v$ is expected to be strongly
renormalized close to the threshold as explained in Appendix
\ref{sec:appendix_vortex}.).  The 5th, 6th and 7th rows show the values of
$L_{\mathrm{KPZ}}$, $\xi_0$ and $F_{\mathit{lu}}^2$ at this point.  We obtain
that only for the \textit{bad} and \textit{typical} configurations, this
intersection point can be distinguished from the upper threshold.
The last row shows the magnitude of the drift term of the aKPZ
equation in dimensionless units for $k_p=1.4$.  The organic cavity has the
highest values, whereas the typical configuration the lowest one. 
We can conclude that in the state-of-the-art microcavities near the OPO
  threshold the length scales associated with the KPZ or vortex dominated phases
  are absolutely unrealistic and the physics is dominated by the
  equilibrium-like BKT transition between disorder and algebraically ordered
  phases.

Finally, we examine whether the strong KPZ non-linearity, which would result
  in stretched exponential decay of correlations at all lengthscales beyond the
  healing length (cf. Sec.~\ref{sec:finite-syst-search}) is present
  also in other cavity configurations. We find that for longer lifetime cavities
  this regime moves to higher pump powers above the OPO threshold with respect
  to the {\it{bad}} cavity configuration. However, for all cavities other then
  the {\it{bad}} cavity this regime falls into a region, where the three mode
  ansatz is unstable towards more complex solutions. Examining whether the KPZ
  scaling persists beyond the three mode ansatz is beyond the scope of this
  work. 

\begin{table}
\begin{tabular}{c|c|c|c|c|}
Name & \textit{bad} & Typical & Good & Organic \\
\hline
\hline
$L_v(\mu \mathrm{m})$ & $10^2$ & $10^8$ & $10^{27}$ & $2\cdot10^2$ \\
\hline
$L_{\mathrm{KPZ}}(\mu \mathrm{m})$ & $10^{10^3}$ & $10^{10^4}$ & $10^{10^5}$ & $10^{10^8}$ \\
\hline
$\xi_0(\mu \mathrm{m})$ & $10$ & $10$ & $10$ & $1$ \\
\hline
$F_{\mathit{np}}^2$ & $0.9$ & $0.84$ & $0.97$ & $0.84$ \\
\hline
\hline
$L_{\mathrm{KPZ}}(\mu \mathrm{m})$ & $10^{3}$ & $10^{10}$ &  & \\
\hline
$\xi_0(\mu \mathrm{m})$ & $60$ & $10^2$  &     &  \\
\hline
$F_{\mathit{pu}}^2$ & $0.9988$ & $0.9999$ & $1.0000$ & $1.000$ \\
\hline
\hline
$B_d (\mu \mathrm{m}/\mathrm{ps})$ & $0.30$ & $0.18$ & $0.30$ & $2.69$ \\
\hline
\end{tabular}
\caption{\textbf{Comparison of different experimental microcavities}. Each
  column contains details for one of the four different considered
  cavities: \textit{bad, typical, good} and organic
  samples for $k_p=1.4$ and $k_s=0.1$. The first four rows indicate
  characteristic length scales 
  for the vortex dominated phase ($L_v$), for the KPZ-phase $(L_{\mathrm{KPZ}})$, the
  healing length ($\xi_0$), and normalized pump power ($F_{\mathit{pu}}^2$)
  at the point where $L_v$ reaches the minimum value in the stable
  three-mode OPO region
  close to the upper threshold (region A in Fig.~\ref{fig:mean_field}). The 5th,
  6th and 7th rows show 
  $L_{\mathrm{KPZ}}$, $\xi_0$ and $F_{\mathit{pu}}^2$ where $L_{\mathrm{KPZ}}\approx L_v$,
  close to the upper threshold. The blank spaces for the good
  and the organic samples mean that this crossing point occurs
  practically at the
  upper threshold. 
  The last row shows the values of the drift term $B_d$ in dimensional units,
  which is
  approximately constant for the different values of the
  external drive (see Fig.~\ref{fig:kpz_parameters_galaa} for the
  \textit{bad} cavity case). It is clear that the KPZ lengthscale is unrealistic in regions A and C of the phase diagram presented in Fig. \ref{fig:mean_field}, leaving region B as the only potentially promising regime. }
\label{table:setups}
\end{table}

\begin{figure}
  \includegraphics[width=0.4\textwidth]{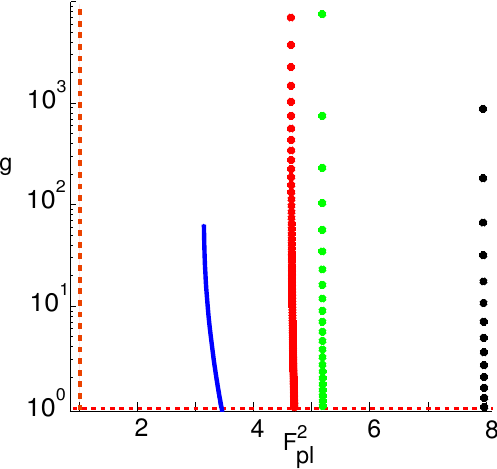}
  \caption { \textbf{Regions potentially exhibiting KPZ phase for
      different experimental microcavities.} KPZ non-linearity
    parameter $g$ (in a region where $g>1$) as a function of the normalized pump strength
    $F^2_{pl}$ for four different microcavities, \emph{bad} (blue
    line), \emph{typical} (red dots), \emph{good} (black dots) and
    organic (green dots) cavities.  We present the regions which
    potentially could show KPZ scaling at all lengthscales beyound
    the healing lenght, i.e. $g>1$ and $\Gamma>1$ (see
    Sec.\ref{sec:infinite_crossover}). The dashed horizontal line
    marks the $g=1$ border, whereas the dashed vertical line indicates the
    normalized lower mean field threshold, i.e. $F^2_{pl} =
    1$. Note that only for the  {\it bad} cavity the three-mode
      ansatz is stable in shown region. In this plot
   we consider $k_p=1.4$ and $k_s=0.1$ for the {\it bad} and \emph{good} cavities and
    $k_p=1.6$ and $k_s=0.06$ for the {\it typical} and organic cavities.  }
\label{fig:kpz_all_cavities}
\end{figure}

\subsection{Summary}

We explored the wealth of scaling regimes accessible with
  coherently driven microcavity polaritons.  The basis of our
  analysis is a long-wavelength effective description of OPO polaritons
  in terms of the compact anisotropic KPZ equation~\eqref{eq:akpz}.
A key point is that while the dynamics of both
  incoherently and coherently driven polaritons can be mapped to the aKPZ
  equation, in the latter case a much wider range of parameters is accessible by
  tuning the pump strength, the exciton-photon detuning, and the pump wave
  vector. Ultimately, the reason for this high versatility of OPO polaritons is
  different physics leading to formally the same long-wavelength
  description. In particular, the strongly fluctuating Goldstone mode is the
  phase of the condensate in the case of incoherent pumping, while it is the
  relative phase of signal and idler modes for polaritons in the OPO regime. The
  crucial merit of this high tunability is that the rich scaling behaviour of the
  compact anisotropic KPZ equation becomes accessible in a single experimental
  platform.

Experimentally, the scaling regimes can be distinguished by measuring the
  spatial and temporal decay of the first order coherence
  function~\eqref{eq:g1}: 
in the WA regime, strongly non-linear
  fluctuations of the phase lead to stretched exponential decay of correlations,
  with exponents in the spatial and temporal ``directions'' given by the KPZ
  roughness and dynamical exponents. However, the non-linearity also leads to
  screening of the interactions of vortices which could result in their
  unbinding and thus preclude the observation of KPZ scaling as discussed in
  Appendix~\ref{sec:appendix_vortex}. Either way, both effects are beyond the
  physics of 2D superfluids in equilibrium and it would be intriguing to see
  them in experiments. They occur beyond length and time scales that are
  exponentially large in the KPZ non-linearity. Thus, their observation is
  greatly facilitated in OPO polaritons, in which this non-linearity can become
  of order one. As the OPO threshold is approached in the WA regime, order on
  shorter scales is destroyed through the usual KT mechanism of vortex
  proliferation.

In the SA regime, the non-linearity is irrelevant, and the effective
  long-wavelength theory becomes linear as in thermal equilibrium. Thus, true
  algebraic order is possible.\footnote{See footnote~\ref{SA_vortices}.} The
  effective renormalized noise level that determines whether the system is in
  the ordered or disordered phase depends in a non-trivial way on the pumping
  strength. In particular, it shows reentrant behaviour: upon increasing the
  driving strength the system can first enter the ordered phase and then leave
  it again.

The most intriguing prospect is thus to cross from the WA to the
  SA regime simply by tuning the pumping strength. Order is then
  established because of a change in the effective degree of
  anisotropy. This is shown to be in principle possible
close to the upper OPO threshold for
  negative detuning 
(i.e. blue curve in
  Fig.~\ref{fig:anisotropic_regime_infinity_system}). Then, as the pumping
  strength is increased, the system crosses from WA to SA, and within the SA
  regime from disordered to ordered and back. Another possibility to enter the
  SA regime close to the lower OPO threshold is to increase the pump
  momentum. This shows that all universal scaling regimes of the compact
  anisotropic KPZ equation are in principle accessible with OPO polaritons --- in a
  sufficiently large system.
  
  The relevant length scales above which asymptotic
  universal behaviour can be observed are determined by the strength of the
  non-linearity in the KPZ equation --- a stronger non-linearity implies shorter
  crossover scales and is thus favourable for experimental observation. The mere
  presence of non-linearity in turn reflects that the system is fundamentally out
  of thermal equilibrium, and its value is enhanced if the dynamics is dominated
  by drive and dissipation, i.e., when polaritons have short lifetimes strong
  pumping is required. Our estimates for the relevant scales to enter universal
  scaling regimes 
thus focus on a \textit{bad} cavity (values of
  parameters are given at the beginning of
  Sec.~\ref{sec:results}).
  We find that while close to the 
  OPO threshold both $L_{\mathrm{KPZ}}$ and $L_v$, beyond which we expect KPZ
  scaling and unbinding of vortices (induced by the screening of interactions
  due to the KPZ non-linearity), respectively, are beyond realistic system
  sizes, there is a very promising regime away from the lower and upper
  thresholds in which non-linear effects are strong and $L_{\mathrm{KPZ}}$ and
  $L_v$ are of the order of the condensate healing length. It would be
  desirable  to explore this regime experimentally or by solving the stochastic
  equations of motion~\eqref{eq:SCGPE} numerically.

Finally, we discuss the crossover from WA to SA in
  finite-size systems, while approaching the upper
  threshold. 
  The most interesting scenario would be to make a transition from a disordered
  to an algebraically ordered phase (up to the size of the system) by tuning the
  effective anisotropy from weak to strong (arrow A in
  Fig.~\ref{fig:strong_anisotropy_finite_lengths}). However, implementing this
  with the \textit{bad} cavity parameters is not feasible in realistic system
  sizes. On the other hand, by optimising the exciton-photon detuning and moving
  to higher pump powers (around four times the lower OPO threshold in
  Fig.~\ref{fig:lengths_intermediate_fp_galaa}) we can cross from WA to SA
  regimes in conditions where the dimensionless KPZ non-linearity is larger then one implying
  KPZ scaling at all lengthscales beyond the healing length. From the
  experimental point of view, this means crossing from a KPZ phase with stretched
  exponential decay of spatial and temporal coherence to an algebraically
  ordered phase by increasing the pump strength in a controlled way.

\section{Conclusions}

Exploring a fruitful example of parametrically driven microcavity
polaritons, we have shown that the underlying order in highly driven
and dissipative conditions heavily depends on the details of the
driving process. In particular, a subtle interplay between dissipation
and spatial inhomogeneity, controlled externally, allows one to move
between different phases with different universal properties.

In our example, polaritons in the OPO regime, despite their intrinsic
driven-dissipative nature and highly non-thermal occupations, can be
driven to a phase, characterised by the equilibrium EW universality
class, and thus become indistinguishable at asymptotic length scales
from an equilibrium system, showing algebraic order and superfluidity
even in the thermodynamic limit. This effect roots in the strong
anisotropy that is feasible in the OPO configuration, and is in a
stark contrast to, for example, incoherently driven polaritons, where
algebraic order and superfluidity can only be a finite size effect,
and will not survive in the thermodynamic limit.  However, in the same
OPO regime but at lower pump powers, the physics is governed by the
non-equilibrium fixed point and KPZ universality class. Again, in
contrast to incoherently pumped polaritons, in the middle region of
the OPO phase diagram the KPZ length scales become small, up to the
order of the healing length, suggesting that the KPZ order in a
quantum system might indeed be observed in experiments on
semiconductor microcavities.
Our findings undoubtedly highlight the importance of the details of
the driving mechanism in establishing the relevant order, applicable
to a wide range of collective light-matter systems, and shine a new
light on the ongoing debate about the nature of the polariton ordered
phase in semiconductor microcavities.

%%%%%%%%%%%%%%%%%%%%%%%%%%%%%%%%%%%%%
%%%%%%%%%%%%%%%%%%%%%%%%%%%%%%%%%%%%
\section{Acknowledgements}
%%%%%%%%%%%%%%%%%%%%%%%%%%%%%%%%%%%%
%%%%%%%%%%%%%%%%%%%%%%%%%%%%%%%%%%%%

We would like to thank E. Altman and L. He for helpful discussions. We
acknowledge support from EPSRC (grants EP/I028900/2 and
EP/K003623/2). L.~S. acknowledges funding from the ERC synergy grant
UQUAM.  S. D. acknowledges funding by the German Research Foundation
(DFG) through the Institutional Strategy of the University of Cologne
within the German Excellence Initiative (ZUK 81), and by the European
Research Council via ERC Grant Agreement n.  647434 (DOQS). This
research was supported in part by the National Science Foundation
under Grant No. NSF PHY11-25915.

\appendix

%%%%%%%%%%%%%%%%%%%%%%%%%%%%%%%%%%%%%%%%%%%%%%%%%%%%%%%%%%%%%%%%%%%%%%%%%
%%%%%%%%%%%%%%%%%%%%%%%%%%%%%%%%%%%%%%%%%%%%%%%%%%%%%%%%%%%%%%%%%%%%%%%%%
\section{Mapping to the KPZ equation and numerical results}
%%%%%%%%%%%%%%%%%%%%%%%%%%%%%%%%%%%%%%%%%%%%%%%%%%%%%%%%%%%%%%%%%%%%%%%%%
%%%%%%%%%%%%%%%%%%%%%%%%%%%%%%%%%%%%%%%%%%%%%%%%%%%%%%%%%%%%%%%%%%%%%%%%%
\label{sec:appendix_kpz}

In this section, we consider in detail the mapping in the long-range limit
between the stochastic equations~\eqref{eq:SCGPE}, describing the dynamics of
the OPO condensate, and the aKPZ equation~\eqref{eq:akpz}.
Firstly, as mentioned in Sec.~\ref{sec:long-wavel-theory}, we introduce the
hydrodynamic phase-amplitude representation \eqref{eq:phase_amplitude} into the
dynamical stochastic equations~\eqref{eq:SCGPE}.
In particular, considering the U(1) symmetry of the system (see
Eq.~\eqref{eq:u1_symmetry}), it is convenient to change from the phase variables
$\{\theta_s,\theta_i \}$ to $\{\theta_+,\theta\}$:
\begin{align}
& \theta_s = \theta + \theta_+, & & \theta  =
  \frac{1}{2}( \theta_s - \theta_i   ), \nonumber \\ 
& \theta_i = -\theta + \theta_+, & & \theta_+ = \frac{1}{2}( \theta_s + \theta_i   ),
\label{eq:change_thetas}
\end{align}
where $\theta$ indicates the phase difference between the signal and the idler
states.
Thus, we get the following set of coupled dynamical equations for the
fluctuations:
\begin{align}
\begin{cases}
&\partial_t\theta + D_s(\theta) + \partial_t(\theta_+ - i
\pi_s/\sqrt{\rho_s})= M_s\phi 
- \tilde{\xi}_s,
\\
&\partial_t\theta + D_i(\theta) + \partial_t(-\theta_+ + i
\pi_i/\sqrt{\rho_i})= M_i\phi 
+ \tilde{\xi}_i,
\\
&\partial_t(\theta_p - i \pi_p/\sqrt{\rho_p})= M_p\phi
- \tilde{\xi}_p,
\end{cases}
\label{eq:pre_kpz}
\end{align}
where $\tilde{\xi}_j = \xi_j/(\sqrt{\rho_j}e^{i\varphi_j})$, and we
have introduced the phase variables $\theta, \theta_+$ in place of
$\theta_s, \theta_i$ according to 
\eqref{eq:change_thetas}.  The phase $\theta$ is kept to all orders
whereas the fluctuations grouped together under $\phi^T=(\pi_s \;
\pi_i \; \pi_p \; \theta_+ \; \theta_p)$ are considered up to linear
order. $M_j$ is the \emph{mass} matrix of the mode $j$ and its
coefficients depend on the mean field parameters of the system. The
operators
$D_s(\theta)=-i\omega_{1s}\nabla\theta 
+i\nabla^T\omega_{2s}\nabla\theta
+\nabla^T\theta\omega_{2s}\nabla\theta$
and 
$D_i(\theta)=-i\omega_{1i}\nabla\theta 
+i\nabla^T\omega_{2i}\nabla\theta
-\nabla^T\theta\omega_{2i}\nabla\theta$
are obtained by considering the expansion of the polariton dispersion
relation around $\bf{k}_s$ and $\bf{k}_i$ (see
Eq.~\eqref{eq:expansion_dispersion}).

Due to the U(1) symmetry of the system Eq.~\eqref{eq:u1_symmetry}, the phase
fluctuations $\theta$ is a gapless mode (there is no 'mass' term for such a
fluctuation), whereas the fluctuations grouped under $\phi$ are all gapped.
Note, that we neglect the following terms containing temporal and spatial
derivatives:
$\pi_j\partial_t\theta_j$,$\nabla\pi_j,\nabla^T\omega_{2j}\nabla\pi_j,\pi_j\nabla\theta_j,\pi_j\nabla^T\omega_{2j}\nabla\theta_j,\pi_j\nabla\theta_j$
and $\pi_j\nabla^T\theta_j\omega_{2j}\nabla\theta_j$, which are small compared
to the mass-like contributions in the long range limit of $\pi_j$. Applying the
same criterion, we also omit the spatial derivatives of $\theta_+$ and
$\theta_p$.
The next step is to consider
the real and imaginary parts of Eq.~\eqref{eq:pre_kpz}, and neglect
the time derivatives of the massive modes, since they are 'slow'
variables~\cite{PhysRevX.5.011017}:
\begin{align}
\begin{pmatrix}
\partial_t\theta + \textrm{Re}[D_s(\theta) + \tilde{\xi}_s] \\
                   \textrm{Im}[D_s(\theta) + \tilde{\xi}_s] \\
\partial_t\theta + \textrm{Re}[D_i(\theta) - \tilde{\xi}_i] \\
                   \textrm{Im}[D_i(\theta) - \tilde{\xi}_i] \\
                   \textrm{Re}[\tilde{\xi}_i]                \\
                   \textrm{Im}[\tilde{\xi}_i] 
\end{pmatrix}
=
\begin{pmatrix}
\textrm{Re}[M_s] \\
\textrm{Im}[M_s] \\
\textrm{Re}[M_i] \\
\textrm{Im}[M_i] \\
\textrm{Re}[M_p] \\
\textrm{Im}[M_p] \\
\end{pmatrix}
\phi.
\label{eq:pre_kpz_v2}
\end{align}
In \eqref{eq:pre_kpz_v2} the massive fluctuations can be eliminated,
i.e. they can be expressed solely in terms of the spatial and time
derivatives of $\theta$, and the noise terms.  By considering the last
five relations of \eqref{eq:pre_kpz_v2}, we can express the gapped
modes as: \begin{equation}
\phi^T=\phi^T(\partial_t\theta,\partial_x\theta,\partial_y\theta,\partial_x^2\theta,\partial_y^2\theta,\tilde{\xi}_s,\tilde{\xi}_i,\tilde{\xi}_p), 
\label{eq:gapped_modes}
\end{equation}
and substitute into the first equation of \eqref{eq:pre_kpz_v2} to
obtain a single dynamical stochastic equation for the gapless $\theta$
variable, which has a form of the KPZ equation with a drift term
 (Eq.~\eqref{eq:akpz}).
The different KPZ coefficients in \eqref{eq:akpz}
read: 
\begin{align}
& D_m = \alpha_1\omega_{2s,mm} + \alpha_2\omega_{2i,mm}, \nonumber \\
& \lambda_m = \alpha_3\omega_{2s,mm} + \alpha_4\omega_{2i,mm}, \nonumber \\
& B_m = \alpha_5\omega_{1s,m} + \alpha_6\omega_{1i,m}, \nonumber \\
& 2\Delta =  (\beta^2_1+\beta^2_2)\gamma_s
  +(\beta^2_3+\beta^2_4)\gamma_i+(\beta^2_5+\beta^2_6)\gamma_p 
\end{align}
for $m=x,y$. The $\alpha$ and $\beta$ coefficients come from
\eqref{eq:gapped_modes} while the $\omega$ coefficients originate from the lower
polariton dispersion \eqref{eq:expansion_dispersion}:
$ \omega_{1j}^T = \left( \omega_{1j,x}, \omega_{1j, y} \right) $ and
\begin{equation}
  \label{eq:1}
  \omega_{2j}=
\begin{pmatrix}
\omega_{2j,xx}  &    0          \\
  0     &     \omega_{2j,yy}    \\
\end{pmatrix}.
\end{equation}

\begin{figure}
  \includegraphics[width=0.22\textwidth]{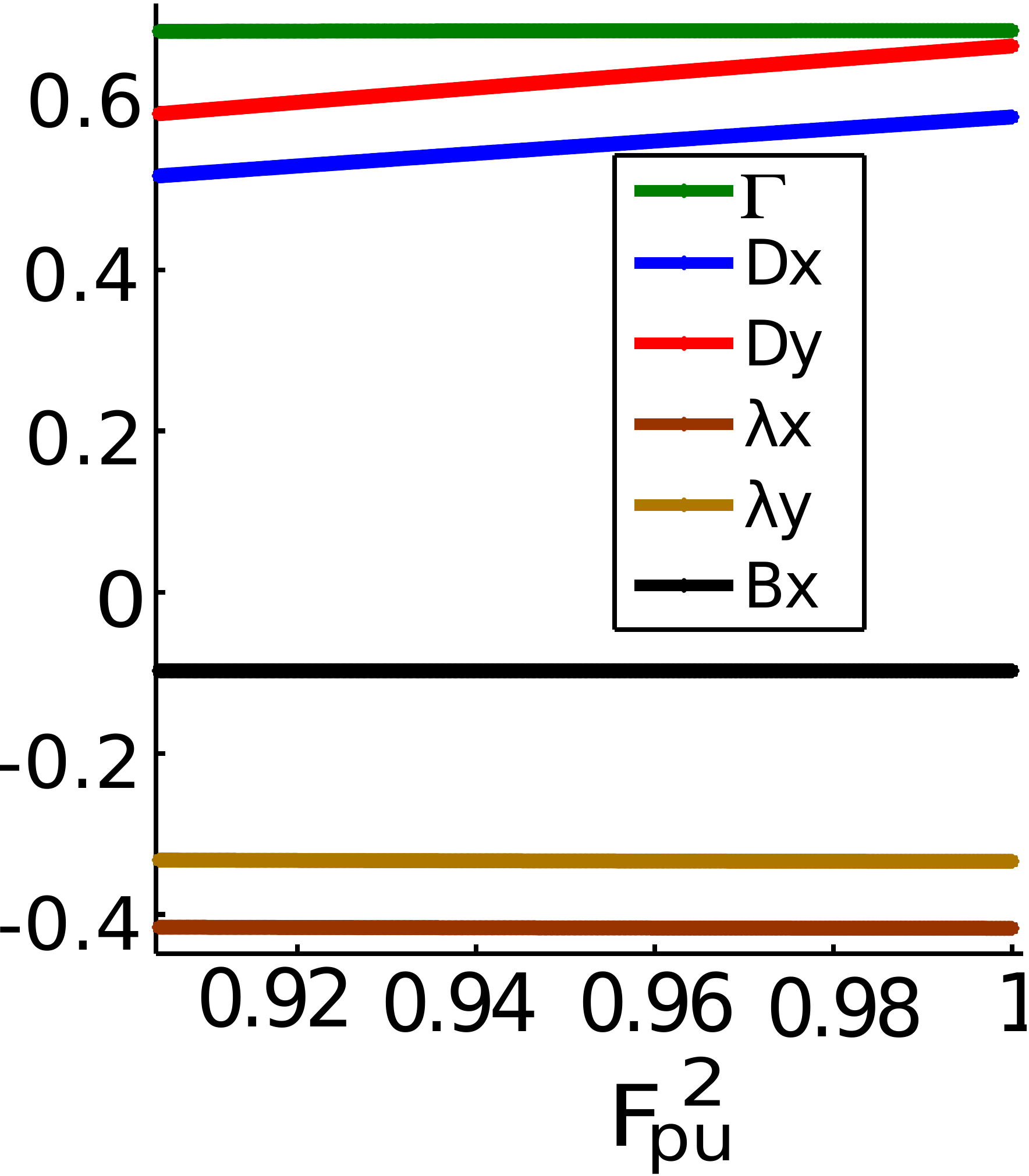}
  \includegraphics[width=0.22\textwidth]{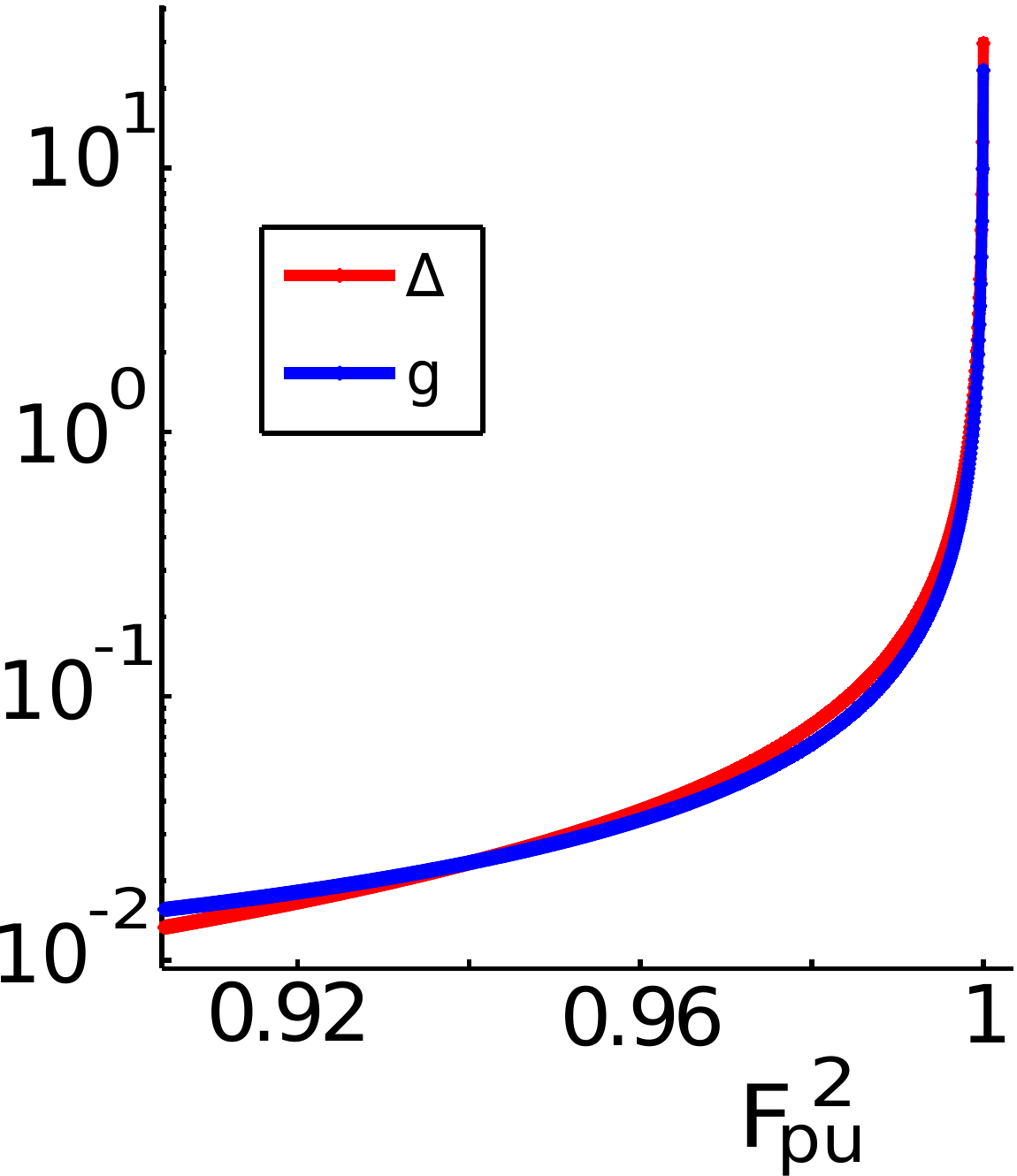}
  \caption { \textbf{aKPZ coefficients for the \textit{bad} cavity configuration with
      zero-detuning.}  \emph{Left panel:} Diffusion coefficients ($D_x,D_y$),
    non-linearities ($\lambda_x,\lambda_y$), anisotropy parameter ($\Gamma$),
    and drift term $B_x$ as a function of the pump power normalised to the upper
    threshold. These parameters vary little with the external pump power.
    \emph{Right panel:} Dimensionless non-linearity $g$ and noise parameter
    $\Delta$ as a function of the normalised pump power. We observe the
    asymptotic growth of these parameters close to the upper OPO threshold.}
\label{fig:kpz_parameters_galaa}
\end{figure}

As an example, in Fig. \ref{fig:kpz_parameters_galaa}, we show the
numerical values of the different KPZ coefficients as a function of
pump power close to the OPO upper threshold and at zero detuning for
the set of parameters given in Sec. \ref{sec:results} characterising
 what we call a \textit{bad} cavity.  The system is WA in the
region shown since $\lambda$s have the same sign. The drift term acts
in the x-direction, due to the choice of the pumping
wave-vector. i.e. $\mathbf{k}_p=(k_p,0)$.  In this WA regime,
the KPZ coefficients do not vary excessively with respect to the
external pump power apart from $g$ and $\Delta$ which asymptote to
infinity at the upper mean-field OPO threshold.

%---- Figure ----------------

%%%%%%%%%%%%%%%%%%%%%%%%%%%%%%%%%%%%%%%%%%%%%%%%%%%%%%%%%%%%%%%%%%%%%%%%%
%%%%%%%%%%%%%%%%%%%%%%%%%%%%%%%%%%%%%%%%%%%%%%%%%%%%%%%%%%%%%%%%%%%%%%%%%
\section{Vortices in the compact KPZ equation in the WA regime}
%%%%%%%%%%%%%%%%%%%%%%%%%%%%%%%%%%%%%%%%%%%%%%%%%%%%%%%%%%%%%%%%%%%%%%%%%
%%%%%%%%%%%%%%%%%%%%%%%%%%%%%%%%%%%%%%%%%%%%%%%%%%%%%%%%%%%%%%%%%%%%%%%%%
\label{sec:appendix_vortex}

In thermal equilibrium, vortices of opposite charge attract each
other with a force that falls of as $\sim 1/r$, like charges in a 2D Coulomb
gas. This leads to the formation of closely bound dipoles at low temperatures,
whereas at high temperatures the interaction is screened at large distances and
the bound state of vortex-antivortex pairs is no longer stable. The fundamental
qualitative modification in a driven-dissipative system is that due to the KPZ
non-linearity the vortex interaction is screened even in the absence of
noise-induced fluctuations~\cite{Aranson1998} (a situation corresponding to zero
temperature in an equilibrium problem). Therefore, even without noise there is
a finite screening length beyond which the interaction is suppressed
exponentially. 
For weak non-linearities, this length scale is given by expression
\eqref{eq:vortex_scale}.

%---- Figure ----------------

\begin{figure}
  \includegraphics[width=0.45\textwidth]{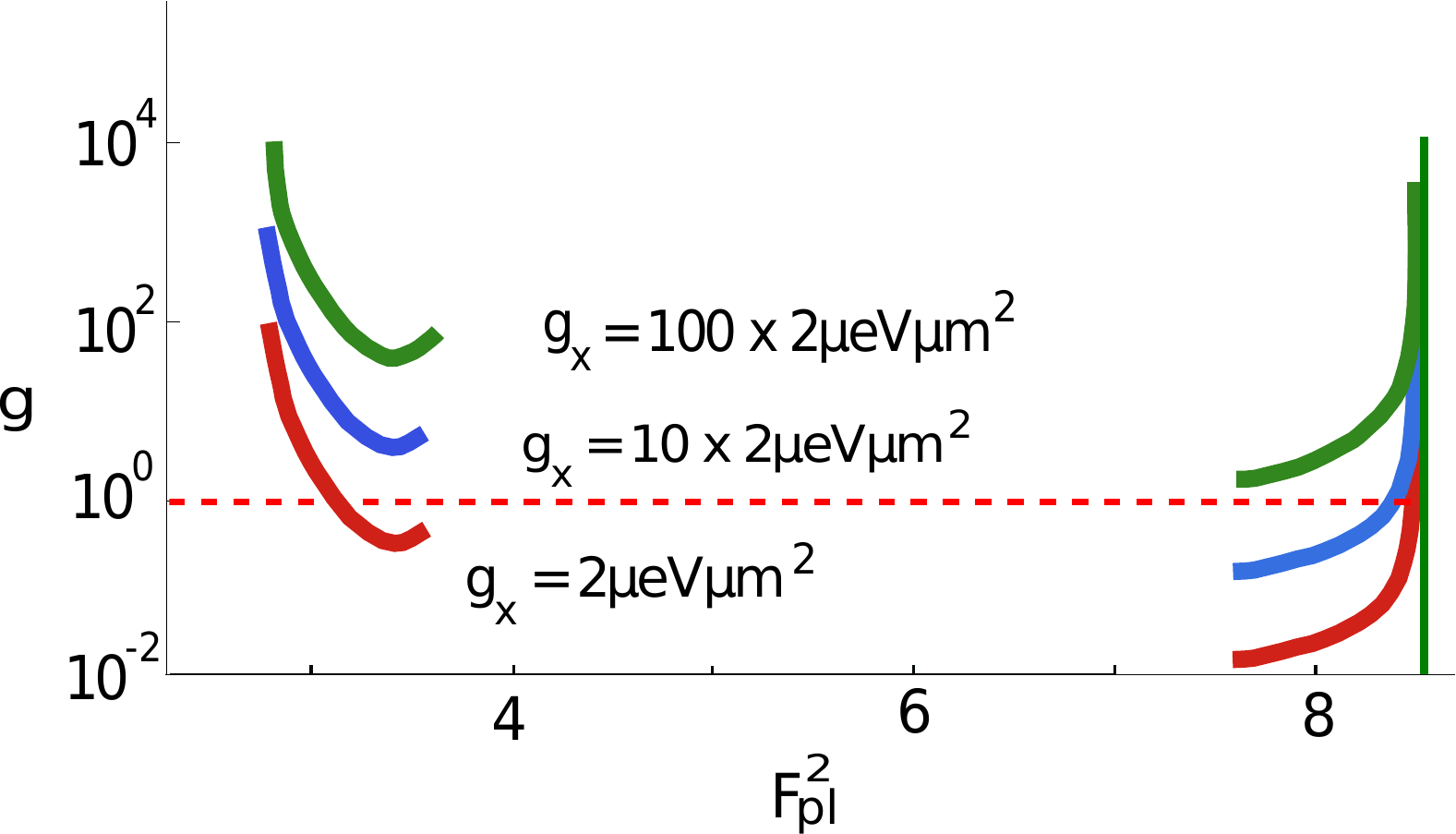}
  \caption { \textbf{Dependence of the non-linear parameter $g$ on
      exciton-exciton interaction strengths.}  
    $g$ for three different values of $g_X$ as a
    function of the normalized external pump power $F_{pl}$ with the lower threshold for a
    cavity with zero detuning. We observe that the systems shows larger values of $g$
    when increasing the exciton-exciton interaction.  }
\label{fig:different_gx}
\end{figure}

As explained above, the screening
of the interaction between vortices beyond the scale $L_v$ is solely due to the
non-linear terms in the KPZ equation. At finite noise, when fluctuations lead to
the creation of vortex-antivortex pairs, there is additional screening induced
by the polarization of bound pairs (which leads to unbinding above the critical
temperature in the usual equilibrium BKT problem). This effect can only be
captured in a proper RG treatment~\cite{wachtel2016electrodynamic} and is not
incorporated in the estimate Eq.~\eqref{eq:vortex_scale}. Associated with $L_v$
is a time scale $t_v$ for vortices to escape the region of attractive
interactions at distances below $L_v$. Then, a vortex-dominated regime
characterized by exponential decay of correlations and in which superfluidity is
destroyed should appear above the scales $L_v$ and $t_v$. If these scales are
smaller than the corresponding KPZ scales defined above, then the scaling
forms~\eqref{eq:kpz_spatial} and~\eqref{eq:kpz_temporal} will be completely
masked by the vortex-induced exponential decay. Indeed, for weak KPZ
non-linearities, some of us estimated that $L_v \ll L_{\mathrm{KPZ}}$ and
$t_v \ll t_{\mathrm{KPZ}}$~\cite{wachtel2016electrodynamic}. The latter estimate
for the time scales relies on the assumption that the mobility of vortices is
not atypically small, i.e., not much smaller than the diffusion coefficients
$D_{x, y}$ in the aKPZ equation~\eqref{eq:akpz}, which are determined by the
same microscopic physics.

Vortex unbinding induced by non-equilibrium conditions has so far remained
elusive in experiments with incoherently pumped, and thus isotropic, polariton
systems, as well as in in the stochastic simulations described
in~\cite{dagvadorj2015nonequilibrium}. This could be ascribed to the limited
length and time scales available to experiments and numerics, but it could also
be taken as an indication that the time scale $t_v$ for vortices to unbind is
indeed much larger than expected, thus leaving open the intriguing possibility
to observe KPZ scaling if the system is initialized in a vortex-free state and
parameters are chosen such that the dimensionless non-linearity $g$ is large
(leading to small values of $t_{\mathrm{KPZ}}$ and $L_{\mathrm{KPZ}}$, see
Eq.~\eqref{eq:length_KPZ}). The most promising regime for observing KPZ physics
is described in Sec.~\ref{sec:finite-syst-search}.

%%%%%%%%%%%%%%%%%%%%%%%%%%%%%%%%%%%%%%%%%%%%%%%%%%%%%%%%%%%%%%%%%%%%%%%%%
%%%%%%%%%%%%%%%%%%%%%%%%%%%%%%%%%%%%%%%%%%%%%%%%%%%%%%%%%%%%%%%%%%%%%%%%%
\section{Effects of the exciton-exciton coupling strength}
%%%%%%%%%%%%%%%%%%%%%%%%%%%%%%%%%%%%%%%%%%%%%%%%%%%%%%%%%%%%%%%%%%%%%%%%%
%%%%%%%%%%%%%%%%%%%%%%%%%%%%%%%%%%%%%%%%%%%%%%%%%%%%%%%%%%%%%%%%%%%%%%%%%
\label{sec:appendix_interaction}

Throughout this paper we considered the exciton-exciton interaction
strength to be $g_X=2 \, \mu \mathrm{eV} \, \mu \mathrm{m}^{-2} $. However,
the true value of $g_X$ is still subject of debates, and different
values have been reported in literature (see for example \cite{walker2015ultra}).  We
consider $g_X=2 \, \mu \mathrm{eV} \, \mu \mathrm{m}^{-2} $ to be the lower
bound, and  the safe upper bound being 100 times this lower value.
Thus, in this section we study the effect of larger values of the
exciton-exciton interactions in two different configurations for the polariton
system, characterized by detunings $\delta_{\mathit{CX}}=0$ and
$\delta_{\mathit{CX}}=-1.08$ in dimensionless units. We consider three different
values for the exciton-exciton interaction:
$g_X=2 \, \mu\textrm{eV} \, \mu \mathrm{m}^{-2}, \; 10\cdot 2 \, \mu\textrm{eV}
\, \mu \mathrm{m}^{-2},\; 100\cdot 2 \, \mu\textrm{eV} \, \mu \mathrm{m}^{-2}$.
In Fig. \ref{fig:different_gx} we show the nonlinear parameter $g$ as a function
of the normalized external pump power (for zero detuning) and as a function of
the anisotropy parameter $\Gamma$ (for finite detuning). We observe that, by
increasing the exciton-exciton interaction constant, the system is characterised
by larger values of the non-linear parameter $g$ for the same values of the
external pump power (see left panel in Fig. \ref{fig:different_gx}). A similar
phenomenon appears for finite detuning, where the value of $g$ also increases by
increasing $g_X$ (see right panel in Fig.~\ref{fig:different_gx}). The
difference is, however, not large enough to alter the conclusions presented in
the main text.

%%%%%%%%%%%%%%%%%%%%%%%%%%%%%%%%%%%%%%%%%%%%%%%%%%%%%%%%%%%
%%%%%%%%%%%%%%%%%  Bibliography  %%%%%%%%%%%%%%%%%%%%%%%%%%
%%%%%%%%%%%%%%%%%%%%%%%%%%%%%%%%%%%%%%%%%%%%%%%%%%%%%%%%%%%

% \bibliography{biblio} % this prints the bibliography section based on the
% \cite commands

%merlin.mbs apsrev4-1.bst 2010-07-25 4.21a (PWD, AO, DPC) hacked
%Control: key (0)
%Control: author (0) dotless jnrlst
%Control: editor formatted (1) identically to author
%Control: production of article title (0) allowed
%Control: page (1) range
%Control: year (0) verbatim
%Control: production of eprint (0) enabled
%

\end{document}